\documentclass[12pt,american,pra,reprint]{revtex4-1}
\usepackage[T1]{fontenc}
\usepackage[latin9]{inputenc}
\usepackage{babel}
\usepackage{mathrsfs}
\usepackage{amsmath}
\usepackage{amssymb}
\usepackage{graphicx}
\usepackage{esint}
\usepackage[unicode=true,pdfusetitle,
 bookmarks=true,bookmarksnumbered=false,bookmarksopen=false,
 breaklinks=false,pdfborder={0 0 1},backref=false,colorlinks=false]
 {hyperref}
\usepackage{breakurl}
\usepackage[raggedright]{subfigure}

\makeatletter
\@ifundefined{textcolor}{}
{%
 \definecolor{BLACK}{gray}{0}
 \definecolor{WHITE}{gray}{1}
 \definecolor{RED}{rgb}{1,0,0}
 \definecolor{GREEN}{rgb}{0,1,0}
 \definecolor{BLUE}{rgb}{0,0,1}
 \definecolor{CYAN}{cmyk}{1,0,0,0}
 \definecolor{MAGENTA}{cmyk}{0,1,0,0}
 \definecolor{YELLOW}{cmyk}{0,0,1,0}
}

\makeatother
\usepackage[matrix,frame,arrow]{xy}
\usepackage{amsmath}

\newcommand{\qw}[1][-1]{\ar @{-} [0,#1]}

\newcommand{\gate}[1]{*{\xy *+<.6em>{#1};p\save+LU;+RU **\dir{-}\restore\save+RU;+RD **\dir{-}\restore\save+RD;+LD **\dir{-}\restore\POS+LD;+LU **\dir{-}\endxy} \qw}

\newcommand{\measureD}[1]{*{\xy*+=+<.5em>{\vphantom{\rule{0em}{.1em}#1}}*\cir{r_l};p\save*!R{#1} \restore\save+UC;+UC-<.5em,0em>*!R{\hphantom{#1}}+L **\dir{-} \restore\save+DC;+DC-<.5em,0em>*!R{\hphantom{#1}}+L **\dir{-} \restore\POS+UC-<.5em,0em>*!R{\hphantom{#1}}+L;+DC-<.5em,0em>*!R{\hphantom{#1}}+L **\dir{-} \endxy} \qw}

\newcommand{\multigate}[2]{*+<1em,.9em>{\hphantom{#2}} \qw \POS[0,0].[#1,0];p !C *{#2},p \save+LU;+RU **\dir{-}\restore\save+RU;+RD **\dir{-}\restore\save+RD;+LD **\dir{-}\restore\save+LD;+LU **\dir{-}\restore}
\newcommand{\ghost}[1]{*+<1em,.9em>{\hphantom{#1}} \qw}
\newcommand{\Qcircuit}[1][0em]{\xymatrix @*=<#1>} 

\newcommand{\pureghost}[1]{*+<1em,.9em>{\hphantom{#1}}}
\newcommand{\multiprepareC}[2]{*+<1em,.9em>{\hphantom{#2}}\save[0,0].[#1,0];p\save !C
  *{#2},p+RU+<0em,0em>;+LU+<+.8em,0em> **\dir{-}\restore\save +RD;+RU **\dir{-}\restore\save
  +RD;+LD+<.8em,0em> **\dir{-} \restore\save +LD+<0em,.8em>;+LU-<0em,.8em> **\dir{-} \restore \POS
  !UL*!UL{\cir<.9em>{u_r}};!DL*!DL{\cir<.9em>{l_u}}\restore}
\newcommand{\prepareC}[1]{*{\xy*+=+<.5em>{\vphantom{#1\rule{0em}{.1em}}}*\cir{l^r};p\save*!L{#1} \restore\save+UC;+UC+<.5em,0em>*!L{\hphantom{#1}}+R **\dir{-} \restore\save+DC;+DC+<.5em,0em>*!L{\hphantom{#1}}+R **\dir{-} \restore\POS+UC+<.5em,0em>*!L{\hphantom{#1}}+R;+DC+<.5em,0em>*!L{\hphantom{#1}}+R **\dir{-} \endxy}}
\newcommand{\poloFantasmaCn}[1]{{{}^{#1}_{\phantom{#1}}}}

\makeatletter

\newcommand{\R}{\mathbb{R}}

\newcommand{\set}[1]{\mathsf{#1}}

\newcommand{\spc}[1]{\mathcal{#1}}

\def\>{\rangle}
\def\<{\langle}

\newcommand{\bs}[1]{\boldsymbol{#1}}     

\newcommand{\map}[1]{\mathcal{#1}}
\newcommand{\Tr}{\operatorname{Tr}}

\newcommand{\Sys}{{\mathsf{Sys}}}
\newcommand{\Rev}{{\mathsf{Rev}}}
\newcommand{\St}{{\mathsf{St}}}
\newcommand{\Eff}{{\mathsf{Eff}}}
\newcommand{\Pur}{{\mathsf{Pur}}}
\newcommand{\Transf}{{\mathsf{Transf}}}

\newtheorem{theo}{Theorem}
\newtheorem{lemma}{Lemma}
\newtheorem{cor}{Corollary}
\newtheorem{defi}{Definition}
\newtheorem{ax}{Axiom}
\newtheorem{prop}{Proposition}

\newtheorem{example}{Example}

\def\Proof{{\bf Proof.~}}
\def\qed{$\blacksquare$ \newline}

\newcommand{\rA}{\mathrm{A}}
\newcommand{\rB}{\mathrm{B}}
\newcommand{\rC}{\mathrm{C}}
\newcommand{\rD}{\mathrm{D}}
\newcommand{\rE}{\mathrm{E}}
\newcommand{\rF}{\mathrm{F}}
\newcommand{\rI}{\mathrm{I}}

\newcommand{\cA}{\mathcal{A}}
\newcommand{\cB}{\mathcal{B}}
\newcommand{\cC}{\mathcal{C}}
\newcommand{\cD}{\mathcal{D}}

\newcommand{\cU}{\mathcal{U}}

\makeatother

  \addto\captionsamerican{}
  \addto\captionsamerican{}
  \addto\captionsamerican{}
  \addto\captionsamerican{}
  \addto\captionsamerican{}
  \addto\captionsamerican{}
  \addto\captionsamerican{}
  \addto\captionsamerican{}
  \addto\captionsbritish{}
  \addto\captionsbritish{}
  \addto\captionsbritish{}
  \addto\captionsbritish{}
  \addto\captionsbritish{}
  \addto\captionsbritish{}
  \addto\captionsbritish{}
  \addto\captionsbritish{}

\begin{document}

\title{Entanglement and thermodynamics  in general probabilistic theories}

\author{Giulio Chiribella}

\email{giulio@cs.hku.hk}

\affiliation{Department of Computer Science, University of Hong Kong, Pokfulam Road, Hong Kong}

\author{Carlo Maria Scandolo}

\email{carlomaria.scandolo@st-annes.ox.ac.uk}

\affiliation{Department of Computer Science, University of Oxford, Oxford, UK}

\begin{abstract}
Entanglement is one of the most striking features of quantum mechanics, and yet it is not specifically quantum. 
More specific to quantum mechanics is the connection  between entanglement and thermodynamics, which leads to an identification between entropies and measures of pure state entanglement. 
 Here we search for the roots of this connection, investigating the relation between entanglement and   thermodynamics in the framework of general probabilistic theories.    
 We first address the question  whether an entangled state   can be transformed into another by means of local operations and classical communication. 
Under two operational requirements, we prove a general version of the  Lo-Popescu theorem,  which lies at the foundations of the theory of pure-state entanglement.  
We then consider a resource theory of purity where free operations are random reversible transformations, modelling the scenario where an agent  has  limited control over the dynamics of a closed system. 
Our key result is a duality between the resource theory of entanglement and the resource theory of purity, valid for every physical  theory  where  all processes arise from pure states and reversible interactions at the fundamental level.     
As an application of the main result,  we establish a one-to-one correspondence between entropies and measures of pure bipartite entanglement.  The correspondence is then used  to define entanglement measures in the general probabilistic framework.  Finally, we show a duality between the task of information erasure and the task of entanglement generation, whereby  the existence of entropy sinks (systems that can absorb arbitrary amounts of information) becomes equivalent to the existence of entanglement sources (correlated systems from which  arbitrary amounts of entanglement can be extracted).  
\end{abstract}


\maketitle

\section{Introduction}

The discovery of quantum entanglement  \cite{EPR,Schrodinger} 
introduced the revolutionary idea that a composite system can be in a pure state while   its components  are  mixed.     In Schr\"odinger's words:  \emph{``maximal knowledge of a total system does not necessarily imply  maximal knowledge of all its parts''} \cite{Schrodinger}. 
This new possibility, 
%
in radical contrast  with the  paradigm of classical physics, 
 is at the root of quantum non-locality \cite{Bell,CHSH,Buhrman,Brunner} in all  its  counterintuitive manifestations    \cite{GHZ,Mermin,Peres,Hardy-paradox1,Hardy-paradox2, Dynamical-nonlocality,Logical-nonlocality}.  With the advent of quantum information, it quickly became clear that entanglement was not only  a source  of foundational puzzles, but also a  resource \cite{Entanglement-resource}.      Harnessing this resource has been the key to the invention of   groundbreaking  protocols such as quantum teleportation   \cite{Teleportation}, dense coding \cite{Dense-coding}, and secure quantum key distribution \cite{QKD,Ekert}, whose implications deeply impacted  physics and computer science  \cite{Mayers-Yao,Vazirani}.  
 
 The key to understand entanglement as a resource is to consider distributed scenarios  where spatially-separated parties perform  local operations (LO) in their laboratories and  exchange  classical communication (CC)  from one laboratory to another \cite{LOCC1,LOCC2,Lo-Popescu}.  
The protocols that can be implemented in this scenario, known as LOCC protocols, provide a means to characterize entangled states and to compare their degree of entanglement. Precisely, a state is  \emph{i)} entangled if it cannot be generated by an LOCC protocol, and \emph{ii)} more entangled than another if there exists an LOCC protocol that transforms the former into the latter.  

 Comparing the degree of entanglement of two quantum states is generally a hard problem \cite{Peres-separable,Horodecki-separable1,Horodecki-separable2,Bound-entanglement1,Bound-entanglement2,W-GHZ,Kraus}.   
Nevertheless, the  solution is simple for  pure bipartite states, where  the  majorization criterion \cite{Nielsen-entanglement}  provides a necessary and sufficient condition for LOCC convertibility.    
 The criterion identifies the degree of entanglement of a  bipartite system with the degree of mixedness of its parts: the more entangled a pure bipartite state is, the more mixed its marginals are.  Mixed states are compared here according to their spectra,  with a state being more mixed than another if the spectrum of the latter majorizes the spectrum of the former  \cite{Uhlmann1,Uhlmann2,Uhlmann3,Geometry-quantum-states}.  
 
 The majorization criterion  shows that  for  pure bipartite states the notion of entanglement as a resource    beautifully matches   Schr\"odinger's  notion of entanglement as non-maximal knowledge about the parts of a pure composite system. Moreover, majorization establishes an intriguing duality  between entanglement and   thermodynamics    
\cite{Popescu-entanglement,Vedral-entanglement,Entanglement-thermodynamics,Brandao,Horodecki-thermodynamics}, whereby   the reduction of entanglement caused by LOCC protocols  becomes dual to the increase of mixedness (and therefore entropy \cite{Thirring}) caused by thermodynamic transformations.  This duality has far-reaching consequences,  such as the existence of a unique measure of pure state entanglement in the asymptotic limit  \cite{Entanglement-concentration,Popescu-entanglement,Vedral-entanglement}. In addition, it has provided guidance for the  development of entanglement theory  beyond the  case of pure bipartite states  \cite{Brandao}.


The duality between entanglement and thermodynamics  is a profound and fundamental fact.  As such, one might expect it to follow  directly from basic principles.   However, 
what these principle are is far from clear:   up to now,  the relation between entanglement and thermodynamics has been addressed in a way that depends  heavily  on the Hilbert space framework, using technical results  that lack an operational interpretation (such as, e.g.\  the singular value decomposition).   
 It is then natural to search for  a derivation of the entanglement-thermodynamics  duality that uses    only high-level  quantum features,  such as the impossibility of instantaneous signalling or the no-cloning theorem. 
 In the same spirit,  one can ask whether the  duality holds for physical theories other than quantum mechanics,  adopting the broad framework of  \emph{general probabilistic theories}    \cite{Hardy-informational-1,Barrett,D'Ariano,Barnum-1,Chiribella-purification,Chiribella-informational,Barnum-2,hardy2011,Hardy-informational-2,Chiribella14}.   
 In the  landscape of general probabilistic theories,  entanglement is a generic feature  \cite{Barrett,barnum2012teleportation}, which  provides powerful advantages for a variety of information-theoretic  tasks    \cite{van1999nonlocality,brassard2006,linden2007,Paw09,almeida2010guess,fritz2013local}.  
But what about its relation with thermodynamics?  Is it also generic, or rather constitutes a specific feature of quantum theory?

In this paper we   explore  the relation between entanglement and thermodynamics in an operational, theory-independent way.  Our work is part of a  larger project that aims at establishing a common axiomatic foundation to  quantum information theory and  quantum thermodynamics. 
Within this broad scope, we start our investigation from the resource theory of entanglement, asking  which conversions are possible under  LOCC protocols. 
   Our first result is a generalization of the Lo-Popescu theorem  \cite{Lo-Popescu}: we  show that under suitable assumptions every LOCC protocol acting on a pure bipartite state can be simulated by a protocol using only one round of classical communication.  Our assumptions are satisfied by quantum theory on both real and complex Hilbert spaces, and also  by   all  bipartite extreme no-signalling boxes  studied in the literature  \cite{PRboxes1,PRboxes0,PRboxes2,PRboxes3}. 
   
 In order to establish the connection  with thermodynamics, we then move our attention to mixed states. 
   We  consider the  scenario where  an agent  has  limited control over the dynamics of a closed system, thus causing it to  undergo  a random mixture of reversible transformations and degrading it to a more disordered state.  This notion of degradation  coincides with the notion of ``adding  noise'' put forward by M\"uller and Masanes for the problem of encoding spatial directions into physical systems \cite{Muller3D}, and represents a natural generalization of the notion of majorization  \cite{olkin}. 
    Provided that that every pure state can be reached from any other pure state through some reversible dynamics, we show that the relevant resource in this scenario  can be identified with the purity of the state.  This observation leads to an operational theory of purity, which in the quantum case turns out to be equivalent to the theory of purity defined by Horodecki and Oppenheim  \cite{Horodecki-Oppenheim}.

     Once the resource theories of entanglement and purity are put into place,   we set out to establish a duality between them.    To this purpose, we consider physical theories that  admit a fundamental level of description where all states are pure and all interactions are reversible. Such theories are identified by the Purification Principle \cite{Chiribella-purification}, which expresses a strengthened version of the  conservation of information  \cite{Chiribella-educational,scandolo14}.   The possibility of a pure and reversilble description  is particularly appealing for the foundations of thermodynamics,  as it reconciles the mixedness of  thermodynamic ensembles with the pure and reversible picture provided by  fundamental physics. 
 In the quantum case, Purification is the starting point for all recent proposals to derive thermodynamic ensembles from the typicality of  pure entangled states \cite{Popescu-Short-Winter,Lubkin,Gemmer-Otte-Mahler,Canonical-typicality,Mahler-book,Concentration-measure,BrandaoQIP2015}, an idea that has been recently explored also in  the broader framework of general probabilistic theories \cite{Dahlsten,Muller-blackhole}. 
   Building on the Purification Principle,   we establish the desired duality between entanglement and thermodynamics, showing that the degree of entanglement of a pure bipartite system coincides with the degree of mixedness of its parts. 
As a consequence, every measure  of single-system mixedness becomes equivalent to a measure of pure bipartite entanglement.  Exploiting this result, we define a class of measures of entanglement, which can be extended to from pure to mixed state via the the convex roof construction  \cite{LOCC2,Vidal,Plenio}, exactly in the same way as in the quantum case.  

Finally, we apply the duality to the task of information erasure  \cite{Landauer}, namely the task of converting a mixed state into a fixed pure state via a set of allowed operations (in our case, the set of random reversible operations).  As a result, erasing information becomes equivalent to generating entanglement. 
 Quite surprisingly, we find out that  the impossibility of erasing information with the assistance of a catalyst implies  the existence of a special type of purification, where the purifying system is a twin of the purified system.    This observation completes the physical picture of the entanglement-thermodynamics duality, which appears to be a consequence of the possibility to describe every physical process in terms of pure states, reversible interactions, and pure measurements---in particular, modelling physical systems in mixed states through the introduction of a mirror image that completes the description.

The paper is organized as follows. In section~\ref{sec:Framework} we introduce the framework.  The resource theory of entanglement is discussed in  section~\ref{sec:Entanglement}.    In section~\ref{sec:Lo-Popescu}  we  prove an operational version of Lo-Popescu theorem, which provides the starting point for the entanglement-thermodynamics duality.  In  section~\ref{sec:Purity} we formulate an operational resource theory of dynamical control, which gives rise to a resource theory of purity under the condition that all pure state are equivalent under the allowed reversible dynamics.  
In section~\ref{sec:Duality} we prove the duality between entanglement and thermodynamics, focussing our attention to  theories that admit a fundamental level where all processes are pure and  reversible. The  consequences of the duality  are examined in section~\ref{sec:consequences}: specifically, we discuss the equivalence between measures of mixedness and measures of entanglement for pure bipartite states, and we establish the relation between  information erasure and entanglement generation. In section~\ref{sec:symmetric purification} we show that the requirement that information cannot be erased for free leads to the requirement of Symmetric Purification.   Finally, section~\ref{sec:Conclusions} draws the conclusions and highlights the  implications of our results.

  

\section{\label{sec:Framework}Framework}

In this paper we adopt the framework of  \emph{operational-probabilistic theories (OPTs)} \cite{Chiribella-purification,Chiribella-informational,Chiribella14}, which combines the toolbox of probability theory with  the graphical language of symmetric monoidal categories \cite{cqm,Coecke-Kindergarten,Coecke-Picturalism,Selinger}.    Here we give a quick recap, referring the reader to Refs.~\cite{Chiribella-purification,Chiribella-informational,Chiribella14} and to Hardy's works \cite{hardy2011,Hardy-informational-2} for a more extended presentation.

The OPT framework describes circuits that can be built up by combining physical processes in sequence and in parallel, as in the following example
\[
\begin{aligned}\Qcircuit @C=1em @R=.7em @!R { & \multiprepareC{1}{\rho}    & \qw \poloFantasmaCn{\rA} &  \gate{\cA} & \qw \poloFantasmaCn{\rA'} &  \gate{\cA'} & \qw \poloFantasmaCn{\rA''} &\measureD{a}   \\  & \pureghost{\rho}    & \qw \poloFantasmaCn{\rB}  &  \gate{\cB} & \qw \poloFantasmaCn{\rB'} &\qw  &\qw &\measureD{b}  }\end{aligned}
\]
Here, $\mathrm{A}$, $\mathrm{A}'$, $\mathrm{A}''$, $\mathrm{B}$,
$\mathrm{B}'$ label physical systems, $\rho$ is a bipartite state,
$\mathcal{A}$, $\mathcal{A}'$ and $\mathcal{B}$ are transformations,
$a$ and $b$ are effects. The two transformations 
 $\mathcal{A}$ and $\mathcal{A}'$ are composed in sequence, while the transformations 
 $\mathcal{A}$ and $\mathcal{B}$ and the effects $a$ and $b$ are composed in parallel.  
 The circuit has no external wires---circuits of this form are associated with probabilities.   
 Two transformations that give the same probabilities in all circuits are identified.   The  short-hand notation $\left(a|\rho\right)$ is used to indicate the probability that the effect $a$ takes place on the state $\rho$, diagrammatically represented as \[
\left(a|\rho\right)~:=\!\!\!\!\begin{aligned}\Qcircuit @C=1em @R=.7em @!R { & \prepareC{\rho}    & \qw \poloFantasmaCn{\rA}  &\measureD{a}}\end{aligned}~.\]

The set of all possible physical systems, denoted by $\Sys$,  is closed under composition:  given two systems $\rA$ and $\rB$ one can form the composite system $\rA\otimes \rB$.   We denote the trivial system as $\rI$, which represents ``nothing''  (or, more precisely, nothing that the theory cares to describe). The trivial system satisfies the obvious conditions $\rA\otimes \rI  =  \rI\otimes \rA  = \rA$, $\forall \rA\in \Sys$.       For generic systems $\rA$ and $\rB$, we denote  as
\begin{itemize}
\item  $\mathsf{St}\left(\mathrm{A}\right)$  the set of states of system $\mathrm{A}$
\item  
$\mathsf{Transf}\left(\mathrm{A},\mathrm{B}\right)$  the set of transformations from system  $\mathrm{A}$ to system  $\mathrm{B}$  
\item  $\mathsf{Eff}\left(\mathrm{A}\right)$ the set of effects on system $\mathrm{A}$ 
\end{itemize}   
The sets of states, transformations, and effects span vector spaces over the real numbers, denoted by $\St_\R  \left(\rA\right)$, $\Transf_\R \left(\rA,\rB\right)$, and $\Eff_\R  \left(\rA\right)$, respectively.  
We denote by $D_\rA$ the dimension of the vector space $\St_\R \left(\rA\right)$ and say that system $\rA$ is \emph{finite} iff  $D_\rA  <+ \infty$.  Transformations and effects act linearly on the vector space of states.    For every system  $\rA$, we assume the existence  of an identity transformation $\map I_\rA$, which does nothing on the states of the system.   

A \emph{test} 
is a collection of transformations 
that can  occur as alternatives  in an experiment.  Specifically, a test of type $\mathrm{A}$ to $\mathrm{B}$ is a collection of transformations $\left\{ \mathcal{C}_{i}\right\} _{i\in \mathsf{X}}$ with input $\mathrm{A}$ and output $\mathrm{B}$. 
 A transformation  is called \emph{deterministic} if it belongs to a test with a single outcome.   We will often refer to deterministic transformations as \emph{channels}, following the standard terminology of  quantum information.   A channel $\mathcal{U}$ from $\rA$ to $\rB$  is called \emph{reversible } if  there exists a channel $\mathcal{U}^{-1}$  from $\rB$ to $\rA$   such that $\mathcal{U}^{-1}\mathcal{U}=\mathcal{I}_{\mathrm{A}}$ and $\mathcal{U}\mathcal{U}^{-1}=\mathcal{I}_{\mathrm{B}}$.  We denote by $\Rev\Transf\left(\rA,  \rB\right)$ the set of reversible transformations from $\rA$ to $\rB$.     If there exists a reversible channel transforming $\rA$ into $\rB$ we say that $\rA$ and $\rB$ are \emph{operationally equivalent}, denoted by $\rA  \simeq  \rB$.

The composition of systems is required to be \emph{symmetric} \cite{cqm,Coecke-Kindergarten,Coecke-Picturalism,Selinger}, meaning that  $\rA\otimes \rB  \simeq  \rB\otimes \rA$.  The reversible channel that implements the equivalence is the \emph{swap channel}, $\tt SWAP$, and satisfies the condition   
\begin{align}\label{swap}
\begin{aligned}\Qcircuit @C=1em @R=.7em @!R {  & \qw \poloFantasmaCn{\rB} &  \gate{ {\cB}} & \qw \poloFantasmaCn{\rB'}  &\qw   \\
   & \qw \poloFantasmaCn{\rA}   &  \gate{ {\cA}} & \qw \poloFantasmaCn{\rA'}  &\qw }
  \end{aligned}   
   = 
\begin{aligned}\Qcircuit @C=1em @R=.7em @!R {  & \qw \poloFantasmaCn{\rB} &    \multigate{1}{{\tt SWAP}}  & \qw \poloFantasmaCn{\rA}        &  \gate{ {\cA}} & \qw \poloFantasmaCn{\rA'}   &             \multigate{1}{{\tt SWAP}}  & \qw \poloFantasmaCn{\rB'}   &  \qw
 \\   & \qw \poloFantasmaCn{\rA}  & \ghost{{\tt SWAP}}  &   \qw \poloFantasmaCn{\rB}     &  \gate{ {\cB}} & \qw \poloFantasmaCn{\rB'}  &  \ghost{{\tt SWAP}}  &   \qw \poloFantasmaCn{\rA'}     &    \qw}
  \end{aligned} ~ ,
\end{align}
for every pair of transformations $\cA$ and $\cB$ and for generic systems $\rA,\rA',\rB,\rB'$, as well as the conditions  
\[
\begin{aligned}
\Qcircuit @C=1em @R=.7em @! R {
& \poloFantasmaCn{\rA} \qw &\multigate{1}{  \tt SWAP }&\poloFantasmaCn{\rB} \qw & \multigate{1}{  \tt SWAP }&\poloFantasmaCn{\rA} \qw &\qw  \\
& \poloFantasmaCn{\rB} \qw &\ghost{  \tt SWAP  }&\poloFantasmaCn{\rA} \qw &\ghost{  \tt SWAP}&\poloFantasmaCn{\rB} \qw &\qw }
\end{aligned}  \  =  ~  
\begin{aligned}
\Qcircuit @C=1em @R=.7em @! R {
&\qw &  \poloFantasmaCn{\rA} \qw &  \qw&\qw \\
&&&&\\
&\qw &  \poloFantasmaCn{\rB} \qw &  \qw&\qw }   
\end{aligned}  ~ ,
\]
the wires in the r.h.s. representing identity transformations, and 
\[
\begin{aligned}
\Qcircuit @C=1em @R=.7em @! R {
& \poloFantasmaCn{\rA} \qw &\multigate{2}{  \tt SWAP }&\poloFantasmaCn{\rB} \qw &\qw\\
& \poloFantasmaCn{\rB} \qw &\ghost{  \tt SWAP }&\poloFantasmaCn{\rC} \qw &\qw\\
& \poloFantasmaCn{\rC} \qw &\ghost{  \tt SWAP  }&\poloFantasmaCn{\rA} \qw &\qw}
 \end{aligned} 
 \  =  ~
\begin{aligned}
\Qcircuit @C=1em @R=.7em @! R {
& \poloFantasmaCn{\rA} \qw &\multigate{1}{  \tt SWAP  }&\poloFantasmaCn{\rB} \qw &\qw&\qw &\qw \\
& \poloFantasmaCn{\rB} \qw &\ghost{  \tt SWAP   }&\poloFantasmaCn{\rA} \qw &   \multigate{1}{  \tt SWAP }  &    \poloFantasmaCn{\rC} \qw &\qw      \\
&  \poloFantasmaCn{\rC} \qw  &  \qw &\qw & \ghost{  \tt SWAP   }& \poloFantasmaCn{\rA} \qw &\qw    }
 \end{aligned}  ~ .
\] 


In this paper we  restrict our attention   to \emph{causal 
theories} \cite{Chiribella-purification}, namely theories where the choice of future  measurement settings does not influence the outcome probability of  present experiments.       Mathematically, causality is equivalent to the fact that for every system $\mathrm{A}$ there is only one deterministic effect, which we denote here by $\mathrm{Tr}_{\mathrm{A}}$, in analogy with the  trace in quantum mechanics.    
The uniqueness of the deterministic effect provides  a canonical way to  define marginal states:
\begin{defi}
The \emph{marginal state} of a bipartite state $\rho_{\mathrm{AB}}$ on system $\rA$ 
is the state   $\rho_{\mathrm{A}}:=\Tr_{\mathrm{B}}\rho_{\mathrm{AB}}$    obtained by applying the deterministic effect on $\mathrm{B}$.
\end{defi}
Moreover, one can define the \emph{norm} of a state $\rho$ as
\[
\left\Vert \rho\right\Vert :=\mathrm{Tr}\,\rho.
\]
The set of normalized states
of $\mathrm{A}$ will be denoted by  \[\mathsf{St}_{1}\left(\mathrm{A}\right)  :   =  \left\{   \rho \in\St \left(\rA\right) \:|\: \left\Vert  \rho  \right\Vert  = 1\right\}  \,.  
\]
In a causal theory, every state is proportional to a normalized state \cite{Chiribella-purification}. In quantum mechanics,  $\St_1  \left(\rA\right)$ is the set of normalized density matrices of system $\rA$, while $\St \left(\rA\right)$ is the set of all sub-normalized density matrices.    
  
In a causal theory channels admit a simple characterization, which will be useful later in the paper:  
\begin{prop}\label{prop:normalization}
Let $\mathcal{C}\in\mathsf{Transf}\left(\mathrm{A},\mathrm{B}\right)$.
$\mathcal{C}$ is a channel if and only if $\mathrm{Tr}_{\mathrm{B}}\mathcal{C}=\mathrm{Tr}_{\mathrm{A}}$.\end{prop}
The proof can be found in lemma 5 of Ref.~\cite{Chiribella-purification}.   



\subsection{Pure states and transformations}

In every probabilistic theory one can define
pure states, and, more generally, pure transformations. Both concepts are based on the notion of \emph{coarse-graining}, i.e.\ the operation of joining two or more outcomes of a test.  
More precisely,  a test $\left\{ \mathcal{C}_{i}\right\} _{i\in \mathsf{X}}$ is a \emph{coarse-graining}
of the test $\left\{ \mathcal{D}_{j}\right\} _{j\in \mathsf{Y}}$ if there
is a partition $\left\{ \mathsf{Y}_{i}\right\} _{i\in \mathsf{X}}$ of $\mathsf{Y}$ such that
$\mathcal{C}_{i}=\sum_{j\in \mathsf{Y}_{i}}\mathcal{D}_{j}$ for every $i\in \mathsf{X}$.
In this case, we say that $\left\{ \mathcal{D}_{j}\right\} _{j\in \mathsf{Y}}$
is a \emph{refinement} of $\left\{ \mathcal{C}_{i}\right\} _{i\in \mathsf{X}}$.  The refinement of a given transformation is defined via the refinement of a test:      if $\left\{\map D_j\right\}_{j\in\set Y}$ is a refinement of $\left\{\map C_i\right\}_{i\in\set X}$, then the transformations $\left\{\map D_j\right\}_{j\in\set Y_i}$ are a refinement of the transformation $\map C_i$.  


A transformation is called \emph{pure} if it has only trivial refinements:  
\begin{defi}\label{def:puredecomp}
The transformation $\mathcal{C}  \in  \Transf\left(\rA, \rB\right)$   is \emph{pure} if for every refinement $\left\{\map D_j\right\}$ one has $  \map D_j  =   p_j  \map C$, where $\left\{  p_j\right\}$ is a probability distribution. 
\end{defi}
Pure transformations are those for which the experimenter has maximal information about the evolution of the system.    We assume as part of the framework that tests satisfy a \emph{pure decomposition property}: 
\begin{defi}
A  test  satisfies the \emph{pure decomposition property} if it admits a refinement consisting only of pure transformations.  
\end{defi}
Later in the paper, we will assume one axiom---Purification---that implies the validity of the pure decomposition property for every possible test \cite{Chiribella-purification}.  

The set of pure transformations from $\rA$ to $\rB$  will be denoted as  $\Pur\Transf\left(\rA, \rB\right)$.       In the special case of states (transformations with no input), the above definition coincides with the usual definition of pure state.   
We denote the set of pure states of system $\mathrm{A}$ as $\mathsf{PurSt}\left(\mathrm{A}\right)$. As usual, non-pure states will be called \emph{mixed}.

Pure states  will play a key role in this paper. An elementary property of pure states is that they are preserved by reversible transformations.        
\begin{prop}
\label{prop:pure states reversible}  
Let $\mathcal{U}\in\mathsf{Transf}\left(\mathrm{A},\mathrm{B}\right)$ be a reversible
channel. Then a state $\psi  \in  \St \left(\rA\right)$ is pure if and only if the state $\mathcal{U}\psi  \in\St \left(\rB\right)$
is pure.\end{prop}
The proof is standard and is reported in  Appendix~\ref{app:proof_reversible} for convenience of the reader.  

\section{\label{sec:Entanglement}The resource theory of entanglement}

The resource theory of quantum entanglement  \cite{Entanglement-resource} is based on the notion of LOCC protocols, that is, protocols in which distant parties are allowed to communicate classically to one another and to perform local operations in their laboratories  \cite{LOCC1,LOCC2}.    Being operational, the notion of LOCC protocol can be directly exported to arbitrary theories. 

In this paper we  consider protocols involving only  two parties, Alice and Bob.  A generic LOCC protocol  consists of a sequence of tests, performed  by Alice and Bob, with the property that the choice of  the test at a given step can depend on all the outcomes produced at the previous steps.  
For example,  consider a  two-way protocol where
\begin{enumerate}
\item Alice performs a test $\left\{\map A_{i_{1}}\right\}$ and communicates the outcome to Bob
\item Bob performs a test $\left\{ \map B^{\left(i_{1}\right)}_{i_{2}} \right\}$ and communicates the outcome to Alice
\item Alice performs a test  $\left\{\map A^{\left(i_{1},i_{2}\right)}_{i_{3}}\right\}$.
\end{enumerate}
An  instance of the protocol is identified by the sequence of outcomes $\left(i_1,i_2,i_3\right)$ and  can be represented by a circuit of the form
\[
\begin{aligned}
\Qcircuit @C=1em @R=.7em @!R 
{  & \qw \poloFantasmaCn{\rA_0} &  \gate{ \cA_{i_{1}}  }\ar@{-->}[drr] & \qw &  \qw \poloFantasmaCn{\rA_1}&\qw &  \gate{\cA^{(i_{1},i_2)}_{i_3} }   &  \qw  \poloFantasmaCn{\rA_2}  &\qw  \\  
    & \qw \poloFantasmaCn{\rB_0}  & \qw &\qw  &  \gate{ \cB^{(i_{1})}_{i_{2}} }\ar@{-->}[urr] & \qw &\qw   \poloFantasmaCn{\rB_1}   & \qw &\qw  }\end{aligned} ~,
\]
where the dashed arrows represent classical communication.  By coarse-graining over  all possible outcomes, one obtains a channel, given by 
\[
\map L     =      \sum_{i_{1},i_{2},i_{3}}     \left[ \map A^{\left(i_{1},i_{2}\right)}_{i_{3}}  \map A_{i_{1}}  \otimes \map B_{i_{2}}^{\left(i_{1}\right)}   \right] \, .
\]


Entangled states are  those states  that cannot be generated  using an LOCC protocol.    
 Equivalently, they can be characterized as the states that are not \emph{separable}, i.e.\ not of the form 
  \[
  \rho   =  \sum_i  p_i       \,   \alpha^{\left(i\right)}\otimes   \beta^{\left(i\right)} \, , 
  \]
  where $\left\{p_i\right\}$ is a probability distribution allowed by the theory,  $\alpha^{\left(i\right)}$ is a state of $\rA$, and $\beta^{\left(i\right)}$ is a state of $\rB$.   
  
Like in quantum theory, LOCC protocols can be used 
 to compare entangled states.     
 \begin{defi}
 Given two states $\rho  \in   \St \left(\rA\otimes \rB\right)$ and $\rho'  \in \St \left(\rA'\otimes \rB'\right)$, we say that $\rho$ is \emph{more entangled}  than  $\rho'$, denoted by $\rho  \succeq_{\rm ent}  \rho'$, if there exists an LOCC protocol that transforms $\rho$ into $\rho'$, i.e.\ if   $\rho'   =  \map L  \rho$ for some LOCC channel $\map L$.    
\end{defi}  
Mathematically, the relation $\succeq_{\rm ent}$ is a preorder, i.e.\ it is reflexive and transitive.   Moreover, it is stable under tensor products, namely $\rho\otimes \sigma  \succeq_{\rm ent}  \rho'  \otimes \sigma'$ whenever $\rho   \succeq_{\rm ent}   \rho'$ and $\sigma \succeq_{\rm ent}    \sigma'$.    In other words, the relation $\succeq_{\rm ent}$  turns the set of all bipartite states into a preordered monoid, the typical mathematical structure arising in all resource theories \cite{Spekkens}.     The resource theory of entanglement here fits completely  into the framework of Ref.~\cite{Spekkens}, with the LOCC channels as free operations.  The states which can be prepared by LOCC (i.e.\ the separable states) are free by definition, and all the other states represent resources.  
  If $\rho \succeq_{\rm ent} \rho'$  and $\rho'  \succeq_{\rm ent}  \rho$, then we say that $\rho$ and $\rho'$ are \emph{equally entangled}, denoted by $\rho   \simeq_{\rm ent}   \rho'$.   Note that $\rho   \simeq_{\rm ent}   \rho'$  does not imply that  $\rho$ and $\rho'$ are equal:  for example, every two separable states are equally (un)entangled. 

\section{\label{sec:Lo-Popescu}An operational Lo-Popescu theorem}

Given two bipartite states, it is natural to ask whether one is more entangled than the other.   A priori, answering the question requires one to check all possible LOCC protocols. 
However, the situation is much simpler when the initial state is pure:  here we prove that in this case every LOCC protocol can be replaced without loss of generality by a protocol involving only one round of classical communication---i.e.\ a \emph{one-way LOCC protocol}.  Our argument is based on two basic operational requirements and provides a generalized version of the Lo-Popescu theorem \cite{Lo-Popescu}, the key result at  the foundation of the quantum theory of pure-state entanglement.

\subsection{Two operational requirements}
Our derivation of  the operational Lo-Popescu theorem is based on two requirements, the first being 
\begin{ax}[Purity Preservation \cite{D'Ariano,Chiribella-informational,scandolo14}]\label{ax:purpres}  
The composition of two pure transformations yields a pure transformation, namely   
\begin{align*}
&\begin{aligned}
\Qcircuit @C=1em @R=.7em @! R {
& \poloFantasmaCn{\rA} \qw &\multigate{1}{  \map A  }&\poloFantasmaCn{\rA'} \qw &\qw \\
&     &\pureghost{  \map A  }&\poloFantasmaCn{\rC} \qw &\qw    }
 \end{aligned}       \quad   {\rm pure} \, ,   \quad  
     \begin{aligned}
\Qcircuit @C=1em @R=.7em @! R {
&    \poloFantasmaCn{\rC} \qw   &\ghost{  \map B  }& & \\    
& \poloFantasmaCn{\rB} \qw &\multigate{-1}{  \map B  }&\poloFantasmaCn{\rB'} \qw &\qw \\
}
 \end{aligned}       \quad   {\rm pure}  
   \\  \\ 
&\Longrightarrow \quad
\begin{aligned}
\Qcircuit @C=1em @R=.7em @! R {
& \poloFantasmaCn{\rA} \qw &\multigate{1}{  \map A  }&\poloFantasmaCn{\rA'} \qw &\qw&\qw &\qw \\
&   &\pureghost{ \map A    }&\poloFantasmaCn{\rC} \qw &   \multigate{1}{  \map B }  &    &        \\
&  \poloFantasmaCn{\rB} \qw  &  \qw &\qw & \ghost{  \map B  }& \poloFantasmaCn{\rB'} \qw &\qw    }
 \end{aligned}    
 \quad  {\rm pure}  \, ,
\end{align*} 
for every choice of systems $\rA,\rA',\rB,\rB',\rC$.
\end{ax}
As a special case, Purity Preservation implies that 
the product of two pure states is a pure state.
This conclusion could also be obtained  from the Local Tomography axiom \cite{Hardy-informational-1,Barrett,D'Ariano,Chiribella-purification}. Nevertheless,   counterexamples  exist of theories that satisfy Purity Preservation and violate Local Tomography. An  example is quantum theory on real vector spaces  \cite{stueckelberg,wootters1990local,hardy2012limited}. In general, we regard Purity Preservation as  more fundamental than Local Tomography.  Considering the theory as an algorithm to make deductions about physical processes, Purity Preservation ensures that, when presented with maximal information about two processes, the algorithm outputs maximal information about their composition \cite{scandolo14}.   

Our second requirement imposes a symmetry of pure bipartite states:
  \begin{ax}[Local Exchangeability]
\label{ax:locex} For every  pure bipartite  state $\Psi\in \Pur\St \left(\rA\otimes \rB\right) $,
there exist two channels $\mathcal{C}\in\mathsf{Transf}\left(\mathrm{A},\mathrm{B}\right)$
and $\mathcal{D}\in\mathsf{Transf}\left(\mathrm{B},\mathrm{A}\right)$
such that  
\begin{align}\label{locswap}
\begin{aligned}\Qcircuit @C=1em @R=.7em @!R { & \multiprepareC{1}{\Psi}    & \qw \poloFantasmaCn{\rA} &\gate{\cC} & \qw \poloFantasmaCn{\rB} &\qw \\  & \pureghost{\Psi}    & \qw \poloFantasmaCn{\rB}  &\gate{\cD} & \qw \poloFantasmaCn{\rA} &\qw}\end{aligned} ~=\!\!\!\! \begin{aligned}\Qcircuit @C=1em @R=.7em @!R { & \multiprepareC{1}{\Psi}    & \qw \poloFantasmaCn{\rA} &   \multigate{1}{  {\tt  SWAP}}   &  \qw  \poloFantasmaCn{\rB}  &  \qw   \\  & \pureghost{\Psi}    & \qw \poloFantasmaCn{\rB}   &   \ghost{  {\tt  SWAP}}   &    \qw \poloFantasmaCn{\rA} &\qw  } \end{aligned}    ~ .
\end{align}
where $\tt SWAP$ is the swap operation [cf.\ Eq.~\eqref{swap}].   
\end{ax}
Note that, in general, the two channels depend on the specific pure state $\Psi$.

Local Exchangeability is trivially satisfied by classical probability theory, where all pure states are of the product form. Less trivially, it is satisfied by quantum theory,   both on  complex and on real Hilbert spaces. This fact is illustrated in the following

\begin{example}      Suppose that $\rA$ and $\rB$ are quantum systems, and let $\spc H_{\rA}$ and $\spc H_{\rB}$ be the corresponding Hilbert spaces.  By the Schmidt decomposition, every pure state in the tensor product Hilbert space    can be written as  $$  |\Psi\>   =  \sum_{i=1}^r  \,   \sqrt{p_i}   \,    |\alpha_i\>  \otimes |\beta_i\>  \, ,$$ 
where $\left\{|\alpha_i\>\right\}_{i=1}^r \subset  \spc H_{\rA}$  and $\left\{|\beta_i\>\right\}_{i=1}^r \subset \spc H_{\rB}$ are orthonormal vectors.   The Schmidt decomposition implies the relation 
\begin{align*}
{\tt SWAP}    |\Psi\>    =    \left(C\otimes D\right)  |\Psi\> \, ,  
\end{align*}
where $C$ and $D$ are the partial isometries $C:  =  \sum_{i=1}^r    |\beta_i\>\<\alpha_i| $ and $ D  : =  \sum_{i=1}^r   |\alpha_i\>\<  \beta_i|$.    From the partial isometries $C$ and $D$ it is immediate to construct the desired channels $\map C$ and $\map D$, which can be defined as 
\begin{align*}
\map C \left(\rho\right)     &:  =   C  \rho C^\dag  +  \sqrt{ I_\rA  - C^\dag  C}   \rho  \sqrt{ I_\rA  - C^\dag  C}   \\ 
\map D  \left(\sigma\right)    & :  =   D  \sigma D^\dag  +     \sqrt{I_\rB  - D^\dag  D}    \,  \sigma\,    \sqrt{I_\rB  - D^\dag  D}  \, ,
\end{align*}
where $\rho$ and $\sigma$ are  generic input states of systems $\rA$  and $\rB$, respectively. 
 With this definition, one has  
\begin{align*}
\left(\map C \otimes \map D\right)  \left(  |\Psi\>\< \Psi| \right )   =     {\tt SWAP}     \,  |\Psi\>\< \Psi| \,    {\tt SWAP}   \, ,  
\end{align*}
which is the Hilbert space version of the local exchangeability condition of Eq.~\eqref{locswap}.    
\end{example}

Local Exchangeability is also satisfied by all the extreme bipartite non-local
correlations characterized in the literature \cite{PRboxes1,PRboxes0,PRboxes2,PRboxes3}:
\begin{example}
Consider a scenario where two space-like separated parties, Alice and Bob, perform measurements on a pair of systems, $\rA$ and $\rB$, respectively. 
Let $x$
($y$) be the index labelling Alice's (Bob's) measurement setting and let $a$ ($b$) the index labelling the outcome
of the measurement done by Alice (Bob).  
 In the theory known as box world \cite{Barrett,Boxworld} all no-signalling  probability distributions   $p_{ab|xy}$  are physically realizable and represent states of the composite system $\rA\otimes\rB$. 
 Such probability distributions form
a convex polytope \cite{PRboxes2},  whose extreme points are the pure
states of the theory. 

For $x,y,a,b\in\left\{ 0,1\right\} $ the systems $\rA$ and $\rB$ are operationally equivalent. We will denote by $\map I$ the reversible transformation that converts $\rA$ into $\rB$.  The extreme non-local
correlations have been characterized in  \cite{PRboxes1} and are known to be equal to the standard PR-box correlation \cite{PRboxes0}
\begin{align}\label{standardPR}
p_{ab|xy}=\begin{cases}
\frac{1}{2}       & a+b \equiv  xy   \mod 2\\
0 & \textrm{otherwise}
\end{cases} 
\end{align}
up to exchange of $0$ with $1$ in the local settings of Alice and Bob and in the outcomes of their measurements.  In the circuit picture, these operations are described by local reversible transformations: denoting by $\Phi$ the standard PR-box state, one has that every other pure entangled state   $\Psi  \in\Pur\St\left(\rA\otimes \rB\right)$ is of the form  
\begin{align*}
 \begin{aligned}\Qcircuit @C=1em @R=.7em @!R { & \multiprepareC{1}{\Psi}    & \qw \poloFantasmaCn{\rA} &  \qw   \\  & \pureghost{\Psi}    & \qw \poloFantasmaCn{\rB}   &   \qw  }    \end{aligned} 
 ~ =  \!\!\!\!
\begin{aligned}\Qcircuit @C=1em @R=.7em @!R { & \multiprepareC{1}{\Phi}    & \qw \poloFantasmaCn{\rA} &\gate{\cU} & \qw \poloFantasmaCn{\rA} &\qw \\  & \pureghost{\Phi}    & \qw \poloFantasmaCn{\rB}  &\gate{\map V} & \qw \poloFantasmaCn{\rB} &\qw}\end{aligned} \, , 
\end{align*} 
where $\map U$ and $\map  V$ are reversible transformations.       

To see that Local Exchangeability holds, note that  swapping systems $\rA$ and $\rB$ is equivalent to exchanging $x$ with $y$ and $a$ with $b$.  
   Now, the standard PR-box correlation  of Eq.~\eqref{standardPR} is invariant under exchange $x  \leftrightarrow y$, $a  \leftrightarrow b$, meaning that one has 
  \begin{align*}   \begin{aligned}\Qcircuit @C=1em @R=.7em @!R { & \multiprepareC{1}{\Phi}    & \qw \poloFantasmaCn{\rA} &   \multigate{1}{  {\tt  SWAP}}   &  \qw  \poloFantasmaCn{\rB}  &  \qw   \\  & \pureghost{\Phi}    & \qw \poloFantasmaCn{\rB}   &   \ghost{  {\tt  SWAP}}   &    \qw \poloFantasmaCn{\rA} &\qw  }   \end{aligned}  ~  
   =   \!\!\!\!
  \begin{aligned}\Qcircuit @C=1em @R=.7em @!R { & \multiprepareC{1}{\Phi}    & \qw \poloFantasmaCn{\rA} &  \gate{\map I}   & \qw \poloFantasmaCn{\rB}  &      \qw   \\  & \pureghost{\Phi}    & \qw \poloFantasmaCn{\rB}   &     \gate{\map I^{-1}}   & \qw \poloFantasmaCn{\rA}  &  \qw  }    \end{aligned}   
~ .
\end{align*}
Then it is clear that every pure state of $\rA\otimes \rB$ can be swapped by local operations:  indeed, one has  \begin{align}
 \nonumber \begin{aligned}
 \Qcircuit @C=1em @R=.7em @!R { & \multiprepareC{1}{\Psi}    & \qw \poloFantasmaCn{\rA} &   \multigate{1}{  {\tt  SWAP}}   &  \qw  \poloFantasmaCn{\rB}  &  \qw   \\  & \pureghost{\Psi}    & \qw \poloFantasmaCn{\rB}   &   \ghost{  {\tt  SWAP}}   &    \qw \poloFantasmaCn{\rA} &\qw  }     \end{aligned}
   &~=\!\!\!\!
\begin{aligned}\Qcircuit @C=1em @R=.7em @!R { & \multiprepareC{1}{\Phi}    & \qw \poloFantasmaCn{\rA} &\gate{\cU} & \qw \poloFantasmaCn{\rA} &   \multigate{1}{  {\tt  SWAP}}   &  \qw  \poloFantasmaCn{\rB}  &  \qw   
\\  & \pureghost{\Phi}    & \qw \poloFantasmaCn{\rB}  &\gate{\map V} & \qw \poloFantasmaCn{\rB} &   \ghost{  {\tt  SWAP}}   &    \qw \poloFantasmaCn{\rA} &\qw }
\end{aligned} 
\\   
 \nonumber   &~=\!\!\!\!
\begin{aligned}\Qcircuit @C=1em @R=.7em @!R { 
& \multiprepareC{1}{\Phi}    &    \qw \poloFantasmaCn{\rA} &    \multigate{1}{  {\tt  SWAP}}   &  \qw  \poloFantasmaCn{\rB}  & \gate{\map V} & \qw \poloFantasmaCn{\rB}  &\qw  \\  & \pureghost{\Phi}    & \qw \poloFantasmaCn{\rB}  &    \ghost{  {\tt  SWAP}}   &    \qw \poloFantasmaCn{\rA}  & \gate{\map U} & \qw \poloFantasmaCn{\rA}  &\qw }
\end{aligned} 
\\     
\nonumber  &   ~=\!\!\!\!  \begin{aligned}\Qcircuit @C=1em @R=.7em @!R { & \multiprepareC{1}{\Phi}    & \qw \poloFantasmaCn{\rA} &\gate{ \,~\map I~\,}   &   \qw \poloFantasmaCn{\rB}   & \gate{\map V} & \qw \poloFantasmaCn{\rB} &       \qw   
\\  & \pureghost{\Phi}    & \qw \poloFantasmaCn{\rB}    &\gate{\map I^{-1}}   &   \qw \poloFantasmaCn{\rA}    &\gate{\map U} & \qw \poloFantasmaCn{\rA} &    \qw }
\end{aligned}    \\
\label{aaa}  &~=\!\!\!\! \begin{aligned}\Qcircuit @C=1em @R=.7em @!R { & \multiprepareC{1}{\Psi}    & \qw \poloFantasmaCn{\rA} &\gate{\cC} & \qw \poloFantasmaCn{\rB} &\qw \\  & \pureghost{\Psi}    & \qw \poloFantasmaCn{\rB}  &\gate{\cD} & \qw \poloFantasmaCn{\rA} &\qw}
\end{aligned}  ~ , 
\end{align}     
where $\map C  : =   \map V    \,   \map I   \, \map U^{-1}  $ and $\map D:=  \map U  \, \map I^{-1}  \, \map V^{-1} $.    This proves the Local Exchangeability property for all pure bipartite states in the $2$-setting/$2$-outcome scenario.

The situation is analogous in the case of $2$ settings and arbitrary number of outcomes.  Let us denote the setting by $x,y\in\left\{ 0,1\right\} $ and assume that $a$ can take $d_{\rA}$ values, whereas  $b$ can take $d_{\rB}$ values. Then, all extreme non-local correlations
are characterized in  Ref.~\cite{PRboxes2}.  Up to local reversible transformations, they are labelled by a parameter $k\in\left\{ 2,\ldots,\min\left\{ d_{\rA},d_{\rB}\right\} \right\} $
and they are such that 
\begin{equation}
p_{ab|xy}=\begin{cases}
\frac{1}{k}    &     b-a\equiv xy  \mod k  \\
0 & \textrm{otherwise}
\end{cases}  \, .\label{eq:PRbox}
\end{equation}
Thanks to the local equivalence,   it is enough to prove the validity of  Local Exchangeability for correlations in the standard form of Eq.~\eqref{eq:PRbox}.    We distinguish between the two cases  $xy=0$ and $xy=1$.
For $xy=0$,  swapping $x$ with $y$ and $a$ with $b$ has no
effect on $p_{ab|xy}$.  
For $xy=1$,  by  swapping
$x$ with $y$ and $a$ with $b$, one obtains the probability distribution   
\begin{equation*}
p'_{ab|xy}=\begin{cases}
\frac{1}{k} &   a-b  \equiv 1  \mod k\\
0 & \textrm{otherwise}
\end{cases}
\end{equation*}  
This probability distribution can be obtained from the original one by  relabelling the outputs as $a':=k-a$
and $b':=k-b$.  Such a relabelling corresponds to local reversible operations on $\rA$ and $\rB$.      In other words,    Local Exchangeability holds.

Finally, the last category of extreme non-local correlations characterized in the literature corresponds to the case of arbitrary number of settings and to  $2$-outcome measurements.   In this case,  the  extreme correlations are characterized explicitly in Ref.~\cite{PRboxes3}.   Up to local reversible transformations, the pure states are invariant under swap.
 Hence, the same argument used in Eq.~\eqref{aaa} shows that Local Exchangeability holds.   
\end{example}

\subsection{Inverting the direction of classical communication}  

Purity Preservation and Local Exchangeability have an important consequence. For  one-way protocols acting on a pure input state, the direction of classical communication is irrelevant: every one-way LOCC protocol with communication from Alice to Bob can be replaced by a one-way LOCC protocol with communication from Bob to Alice, as shown by the following.
\begin{lemma}[Inverting CC]
\label{lemma:ABBA}
Let $\Psi $ be a pure state of $\rA\otimes \rB$ and let $\rho'$ be a (possibly mixed) state of $\rA'\otimes \rB'$. Under the validity of axioms~\ref{ax:purpres} and \ref{ax:locex}, the following are equivalent:  
\begin{enumerate}
\item 
 $\Psi$ can be transformed into $\rho'$ by a one-way  LOCC protocol with communication from Alice to Bob 
\item   $\Psi$ can be transformed into $\rho'$ by a one-way  LOCC protocol with communication from  Bob to Alice. 
\end{enumerate}
\end{lemma}   

\Proof   Suppose that $\Psi$ can be transformed into $\rho'$ by  a one-way LOCC protocol with communication from Alice to Bob,  namely 
\begin{align}\label{psirho}
\begin{aligned}
\Qcircuit @C=1em @R=.7em @!R { & \multiprepareC{1}{\rho'}    & \qw \poloFantasmaCn{\rA'} & \qw   \\  & \pureghost{\rho'}    & \qw \poloFantasmaCn{\rB'}  & \qw}\end{aligned}  =  
\sum_{i\in\set X}\!\!\!\!
\begin{aligned}\Qcircuit @C=1em @R=.7em @!R { & \multiprepareC{1}{\Psi}    & \qw \poloFantasmaCn{\rA} &  \gate{ {\cA}_{i}}\ar@{-->}[drr] & \qw \poloFantasmaCn{\rA'} &\qw &\qw &\qw \\  & \pureghost{\Psi}    & \qw \poloFantasmaCn{\rB}  & \qw &\qw  &  \gate{ {\cB}^{(i)}} & \qw \poloFantasmaCn{\rB'}  &\qw}\end{aligned}
\end{align}  
where    $\left\{\map A_i\right\}_{i\in\set X}$ is a test, and, for every outcome $i\in\set X$,  $\map B^{(i)}$ is a channel. Note that one can assume  without loss of generality  that all transformations $\left\{\map A_i\right\}_{i\in\set X}$ are  pure: if the transformations are not pure, we can refine them by the pure decomposition property (cf.\ definition~\ref{def:puredecomp}) and apply the argument to the refined test consisting of pure transformations.  

For every fixed $i\in\set X$,  one has  
\begin{align}
\nonumber
&\begin{aligned}\Qcircuit @C=1em @R=.7em @!R { & \multiprepareC{1}{\Psi}    & \qw \poloFantasmaCn{\rA} &  \gate{ {\cA}_{i}} & \qw \poloFantasmaCn{\rA'} &\qw  \\  & \pureghost{\Psi}    & \qw \poloFantasmaCn{\rB}   &   \qw  & \qw     &\qw}\end{aligned}\\
&  \qquad   =  \!\!\!\!
\begin{aligned}\Qcircuit @C=1em @R=.7em @!R { & \multiprepareC{1}{\Psi}    & \qw \poloFantasmaCn{\rA} &    \multigate{1}{{\tt SWAP}}  & \qw \poloFantasmaCn{\rB}      &   \qw &    \qw    &           \multigate{1}{{\tt SWAP}}  & \qw \poloFantasmaCn{\rA'}   &  \qw
 \\  & \pureghost{\Psi}    & \qw \poloFantasmaCn{\rB}  & \ghost{{\tt SWAP}}  &   \qw \poloFantasmaCn{\rA}     &  \gate{ {\cA}_i} & \qw \poloFantasmaCn{\rA'}  &  \ghost{{\tt SWAP}}  &   \qw \poloFantasmaCn{\rB}     &    \qw}
  \end{aligned}
\label{swapswap}
\end{align}
By Local Exchangeability, the first swap can be realized by two local channels $\map C:  \rA\to \rB$ and $\map D: \rB\to \rA$.  
Moreover,   since $\map A_i$ is pure, Purity Preservation implies that the (unnormalized) state $\left(  \map A_i  \otimes \map   I_\rB  \right)   \Psi$ is pure.  Hence, also the second swap in Eq.~\eqref{swapswap} can be realized by two local channels $\map C^{(i)}: \rA'\to \rB  $ and $\map D^{(i)}:  \rB\to \rA'$.  
Substituting into Eq.~\eqref{swapswap} one obtains
\begin{align*}
\nonumber
&\begin{aligned}\Qcircuit @C=1em @R=.7em @!R { & \multiprepareC{1}{\Psi}    & \qw \poloFantasmaCn{\rA} &  \gate{ {\cA}_{i}} & \qw \poloFantasmaCn{\rA'} &\qw  \\  & \pureghost{\Psi}    & \qw \poloFantasmaCn{\rB}   &   \qw  & \qw     &\qw}\end{aligned}\\
& \qquad \qquad \qquad   
= \!\!\!\! \begin{aligned}\Qcircuit @C=1em @R=.7em @!R { & \multiprepareC{1}{\Psi}    & \qw \poloFantasmaCn{\rA}  &\gate{\cC}  &\qw \poloFantasmaCn{\rB} &   \qw  &\qw  &\gate{\cD^{(i)}} & \qw \poloFantasmaCn{\rA'} &\qw  \\  & \pureghost{\Psi}    & \qw \poloFantasmaCn{\rB}  & \gate{\cD} &\qw  \poloFantasmaCn{\rA}  & \gate{\cA_i} &\qw \poloFantasmaCn{\rA'} &\gate{\cC^{(i)}} &\qw \poloFantasmaCn{\rB} &\qw}
\end{aligned} 
\end{align*}
and, therefore, 
\begin{align}
\nonumber
&  \begin{aligned}\Qcircuit @C=1em @R=.7em @!R { & \multiprepareC{1}{\Psi}    & \qw \poloFantasmaCn{\rA} &  \gate{ {\cA}_{i}}\ar@{-->}[drr] & \qw \poloFantasmaCn{\rA'} &\qw &\qw &\qw \\  & \pureghost{\Psi}    & \qw \poloFantasmaCn{\rB}  & \qw &\qw  &  \gate{ {\cB}^{(i)}} & \qw \poloFantasmaCn{\rB'}  &\qw}\end{aligned}  
\\
\nonumber &\\  
\nonumber & \qquad    
~= \!\!\!\! \begin{aligned}\Qcircuit @C=1em @R=.7em @!R { & \multiprepareC{1}{\Psi}    & \qw \poloFantasmaCn{\rA}  &\gate{\cC}  &\qw \poloFantasmaCn{\rB} &   \qw  &\qw  &\gate{\cD^{(i)}} & \qw \poloFantasmaCn{\rA'} &\qw  &\qw &\qw  \\  & \pureghost{\Psi}    & \qw \poloFantasmaCn{\rB}  & \gate{\cD} &\qw  \poloFantasmaCn{\rA}  & \gate{\cA_i}   \ar@{-->}[urr] &\qw \poloFantasmaCn{\rA'} &\gate{\cC^{(i)}} &\qw \poloFantasmaCn{\rB} &\gate{\map B^{(i)}}  &  \qw  \poloFantasmaCn{\rB'}    &\qw}
\end{aligned} \\
\nonumber &  \\
&\qquad ~=: \!\!\!\!  \begin{aligned} 
\Qcircuit @C=1em @R=.7em @!R { & \multiprepareC{1}{\Psi}    & \qw \poloFantasmaCn{\rA}  & \qw  &\qw &\gate{\widetilde{\cA}^{(i)}} &\qw \poloFantasmaCn{\rA'}  & \qw \\  & \pureghost{\Psi}    & \qw \poloFantasmaCn{\rB}  & \gate{\widetilde{\cB}_{i}}\ar@{-->}[urr] &\qw  \poloFantasmaCn{\rB'}  &\qw &\qw &\qw}\end{aligned}  ~ ,
\label{tobesummed}
\end{align}
having defined $  \widetilde{\map A}^{\left(i\right)}  :  =    \map D^{\left(i\right)} \, \map C $ and      $    \widetilde{\map B}^{\left(i\right)}   =   \map B^{\left(i\right)} \, \map C^{\left(i\right)}  \, \map A_i  \,  \map D  $.   By construction $\left\{    \widetilde {\map B}_{i} \right\}_{i\in\set X}$ is a test, because it can be realized by performing the test $\left\{  \map A_i\right\}_{i\in\set X}$  after the channel $\map D$ and subsequently applying the channel $  \map B^{\left(i\right)} \, \map C^{\left(i\right)}$, depending on the outcome (the ability to perform conditional operations  is guaranteed by causality \cite{Chiribella-purification}).       On the other hand, $\widetilde{\map A}^{\left(i\right)}$ is a channel for every $i\in\set X$.  Hence, we have constructed a one-way LOCC protocol with communication from Bob to Alice.  Combining Eqs.~\eqref{psirho} and  \eqref{tobesummed} we  obtain  
\begin{align*}
\begin{aligned}
\Qcircuit @C=1em @R=.7em @!R { & \multiprepareC{1}{\rho'}    & \qw \poloFantasmaCn{\rA'} & \qw   \\  & \pureghost{\rho'}    & \qw \poloFantasmaCn{\rB'}  & \qw}\end{aligned} 
& ~=  
\sum_{i\in\set X} \!\!\!\!
\begin{aligned}\Qcircuit @C=1em @R=.7em @!R { & \multiprepareC{1}{\Psi}    & \qw \poloFantasmaCn{\rA} &  \gate{ {\cA}_{i}}\ar@{-->}[drr] & \qw \poloFantasmaCn{\rA'} &\qw &\qw &\qw \\  & \pureghost{\Psi}    & \qw \poloFantasmaCn{\rB}  & \qw &\qw  &  \gate{ {\cB}^{(i)}} & \qw \poloFantasmaCn{\rB'}  &\qw}\end{aligned}  \\
&  ~=  \sum_{i\in\set X} \!\!\!\!   \begin{aligned} 
\Qcircuit @C=1em @R=.7em @!R { & \multiprepareC{1}{\Psi}    & \qw \poloFantasmaCn{\rA}  & \qw  &\qw &\gate{\widetilde{\cA}^{(i)}} &\qw \poloFantasmaCn{\rA'}  & \qw \\  & \pureghost{\Psi}    & \qw \poloFantasmaCn{\rB}  & \gate{\widetilde{\cB}_{i}}\ar@{-->}[urr] &\qw  \poloFantasmaCn{\rB'}  &\qw &\qw &\qw}\end{aligned}     ~,
\end{align*}  
meaning that $\Psi$ can be transformed into $\rho'$ by a one-way LOCC protocol with communication from Bob to Alice.  Clearly,  the same argument can be applied to prove the converse direction.
 \qed

Note that the target state $\rho'$ need not  be pure:  the fact that the direction of classical communication can be exchanged relies  only on the purity of the input state  $\Psi$. 

\subsection{Reduction to one-way  protocols}\label{subject:LoPop}  
We are now ready to derive the operational version of the Lo-Popescu theorem. Our result shows  that the action of an arbitrary LOCC protocol on a pure state can be simulated by a one-way LOCC protocol:

\begin{theo}[Operational Lo-Popescu theorem]
\label{theo:LP}   
Let $\Psi$ be a pure state of $\rA\otimes \rB$ and $\rho' $ be a (possibly mixed) state of $\rA'\otimes \rB'$.   Under the validity of axioms~\ref{ax:purpres} and \ref{ax:locex}, the following are equivalent 
\begin{enumerate}
\item $\Psi$ can be transformed into $\rho'$ by an LOCC protocol
\item $\Psi$ can be transformed into $\rho'$ by a \emph{one-way} LOCC protocol. 
\end{enumerate}
\end{theo}

\Proof  The non-trivial implication is $1\Longrightarrow 2$.  Suppose that $\Psi$
 can be transformed into $\rho'$ by an LOCC protocol with $N$ rounds of classical communication.   Without loss of generality, we assume that Alice starts the protocol and that all transformations occurring in the first $N-1$ rounds are pure.     

Let $s=  \left(i_1,i_2,  \dots,  i_{N-1}\right)$ be the sequence of all  classical outcomes obtained by Alice and Bob up to step  $N-1$,  $p_s$ be the probability of the sequence $s$, and  $\Psi_{s}$ be the pure state after step $N-1$  conditional on the occurrence of $s$.   For concreteness, suppose that the  outcome  $i_{N-1}$ has been generated on Alice's side.  Then,  the rest of the protocol consists in a test $\left\{\map B^{\left(s\right)}_{i_{N}}\right\}$, performed on Bob's side, followed by a channel $\map A^{\left(s,i_N\right)}$ performed on Alice's side.  
By hypothesis, one has 
\begin{align*}
&\begin{aligned}
\Qcircuit @C=1em @R=.7em @!R { & \multiprepareC{1}{\rho'}    & \qw \poloFantasmaCn{\rA'} & \qw   \\  & \pureghost{\rho'}    & \qw \poloFantasmaCn{\rB'}  & \qw}\end{aligned} 
 =  \sum_s  p_s \quad \times \,   \\
&  \qquad  \times \sum_{i_N}   \!\!\!\! \begin{aligned} 
\Qcircuit @C=1em @R=.7em @!R { & \multiprepareC{1}{\Psi_s}    & \qw &\qw \poloFantasmaCn{\rA_{N-1}}  & \qw &\qw  &\qw &\gate{{\cA}^{(s,i_N)}} &\qw \poloFantasmaCn{\rA'}  & \qw \\  & \pureghost{\Psi_s}    &\qw &  \qw \poloFantasmaCn{\rB_{N-1}}  & \qw &\gate{{\cB}^{(s)}_{i_N}}\ar@{-->}[urr] &\qw  \poloFantasmaCn{\rB'}  &\qw &\qw &\qw}\end{aligned}     ~ .
\end{align*}  
Now,  using lemma~\ref{lemma:ABBA} one can invert the direction of the classical communication in the last round, obtaining 
 \begin{align*}
&\begin{aligned}
\Qcircuit @C=1em @R=.7em @!R { & \multiprepareC{1}{\rho'}    & \qw \poloFantasmaCn{\rA'} & \qw   \\  & \pureghost{\rho'}    & \qw \poloFantasmaCn{\rB'}  & \qw}\end{aligned}  =  \sum_s  p_s  \quad \times \,   \\
&  \qquad  \times   
\sum_{i_N} \!\!\!\!
\begin{aligned}\Qcircuit @C=1em @R=.7em @!R { & \multiprepareC{1}{\Psi_s}    & \qw &  \qw \poloFantasmaCn{\rA_{N-1}} & \qw & \gate{ \widetilde{\cA}^{(s)}_{i_N}}\ar@{-->}[drr] & \qw \poloFantasmaCn{\rA'} &\qw &\qw &\qw \\  & \pureghost{\Psi_s}    &\qw & \qw \poloFantasmaCn{\rB_{N-1}}  & \qw & \qw &\qw  &  \gate{ \widetilde{\cB}^{(s,i_N)}} & \qw \poloFantasmaCn{\rB'}  &\qw}\end{aligned}
\end{align*}  
for a suitable test $\left\{  \widetilde{\cA}_{i_N}^{\left(s\right)}\right\}$ and suitable channels $\widetilde {\cB}^{\left(s,i_N\right)}$.  
Now, since both the $\left(N-1\right)$-th and the $N$-th tests are performed by Alice, they  can be merged into a single test, thus reducing  the original LOCC  protocol to an LOCC protocol with $N-1$ rounds. Iterating this argument for $N-1$  times one finally obtains a one-way protocol.    \qed

In quantum theory, the Lo-Popescu theorem provides the foundation for the resource theory of pure-state entanglement.    Having the operational version of this result will be crucial for our study of the relation between entanglement and thermodynamics.   Before entering into that, however, we need to put into place a suitable resource theory of purity, which will  provide the basis for our thermodynamic considerations. 

\section{\label{sec:Purity} The resource theory of Purity}

\subsection{A  resource theory of dynamical control}\label{subsec:dynamical}  

 Consider the scenario where a closed  system  $\rA$ undergoes a reversible dynamics governed by some parameters under the experimenter's control.  For example, system $\rA$ could be a charged particle moving in an electric field, whose intensity and direction can be tuned in order to obtain a desired trajectory.   
In general,   the experimenter may not have full control and   the actual values of the parameters  may fluctuate randomly.   As a result, the evolution of the system will be described by a \emph{Random Reversible  (RaRe)     channel}, that is a channel $\map R$ of the form
\[
\begin{aligned}\Qcircuit @C=1em @R=.7em @!R {    & \qw \poloFantasmaCn{\rA} &  \gate{\map R } & \qw \poloFantasmaCn{\rA} &\qw}\end{aligned}  =  \sum_{i  \in  \set X }  p_i\,   \begin{aligned}\Qcircuit @C=1em @R=.7em @!R {    & \qw \poloFantasmaCn{\rA} &  \gate{\map U^{(i)} } & \qw \poloFantasmaCn{\rA} &\qw}\end{aligned}  ~ ,    
\]
where  $\left\{\map U^{\left(i\right)}\:|\: i\in\set X\right\}$ is a set of  reversible transformations and  $\left\{  p_i\right\}_{i\in\set X}$ is their probability distribution.  Assuming  that the system remains closed during the whole evolution,   RaRe  channels are the most general transformations the experimenter can implement.   

An important  question  in all problems of control is whether a given input state can be driven to a target state  using the allowed dynamics.   With respect to this task, an input state is   more valuable than another if  the set of target states that can be reached from the former contains the set of target states that can be reached from the latter.     In our model, this idea leads to the following definition.
\begin{defi}[More controllable states]
\label{def:mixedness relation}   Given two states  $\rho$ and $\rho'$ of system $\rA$, we say that $\rho$ is \emph{more controllable} than $\rho'$, denoted by $\rho  \succeq  \rho'$,   if $\rho'$ can be  obtained from $\rho$ via a RaRe  channel. 
\end{defi}
This definition appeared independently in an earlier work by M\"uller and Masanes  \cite{Muller3D}, where the authors explored the use of two-level systems as indicators of spatial directions (cf.\ definition 8 in the appendix).  In this paper we propose  to consider it   as the starting point for an axiomatic theory of thermodynamics. 

Definition~\ref{def:mixedness relation}     fits into the general framework of resource theories \cite{Spekkens}, with RaRe  channels playing the role of free operations.  Note that at this level of generality there are no free states:  since  the experimenter can only control the evolution, every state is regarded as a resource.  Physically, this is in agreement with the fact that the input state in a control problem is not chosen by the experimenter---for example, it can be a thermal state or the ground state of an unperturbed Hamiltonian.       
 
As it is always the case in resource theories, the relation $\succeq $ is reflexive and transitive, i.e.\ it is a preorder. Moreover, since the tensor product of two RaRe channels is a RaRe channel, the relation $\succeq$     is stable under tensor products, namely $\rho\otimes \sigma  \succeq  \rho'  \otimes \sigma'$ whenever $\rho   \succeq    \rho'$ and $\sigma \succeq    \sigma'$.  
For finite systems (i.e.\ systems with finite-dimensional state space), M\"uller and Masanes \cite{Muller3D} showed the additional property  
\begin{align}\label{reveq}  
\rho  \succeq    \sigma  \, ,    \quad\sigma \succeq \rho   \quad \Longrightarrow  \quad  \rho   =  \map U  \sigma \, ,
\end{align}
for some reversible transformation $\map U$.  In other words, two states that are equally controllable can only differ by a reversible transformation.  
 
\subsection{From dynamical control to    purity}\label{sub:purity}
 
There is a close relation between the  controllability  of a state  and its  purity.   For example,  a state that is more controllable than a pure state must also be pure. 
\begin{prop}\label{prop:more than pure}
If   $\psi   \in\St\left(\rA\right)$ is a pure state and $\rho \in\St\left(\rA\right)$ is more controllable than $\psi$, then $\rho$ must be pure.  Specifically, $\rho  = \map U  \psi$ for some reversible channel $\map U$.  
\end{prop}
\Proof Since $\psi$ is pure, the condition 
$   \sum_i p_i  \,  \map U^{\left(i\right)}  \rho   =  \psi $   
implies that $   \map U^{\left(i\right)}  \rho  =  \psi $ for every $i$, meaning that $\rho   =   \map V^{\left(i\right)}   \psi$, where $\map V^{\left(i\right)}$ is the inverse of $\map U^{\left(i\right)}$.  Proposition~\ref{prop:pure states reversible} then guarantees  that  $\rho$ is pure.   \qed  

In other words, pure states can  be reached only from pure states. A natural question is whether every state  can be reached from some  pure state.  The answer is positive in quantum theory and in a large class of theories. Nevertheless, counterexamples exist that prevent an easy identification of the resource theory of dynamical control with a ``resource theory of purity''.   This fact is  illustrated in the following

\begin{figure}
\centering
\subfigure[\label{fig:house} Example of state space leading to a  theory of dynamical control that cannot be interpreted as a theory
of purity. Due to the shape of the state space, the reversible transformations can only be  the identity and the reflection around the vertical axis.  The states on the vertical sides are  maximally controllable (and therefore, maximally resourceful) even though they are not pure. ]{
\includegraphics[scale=0.1]{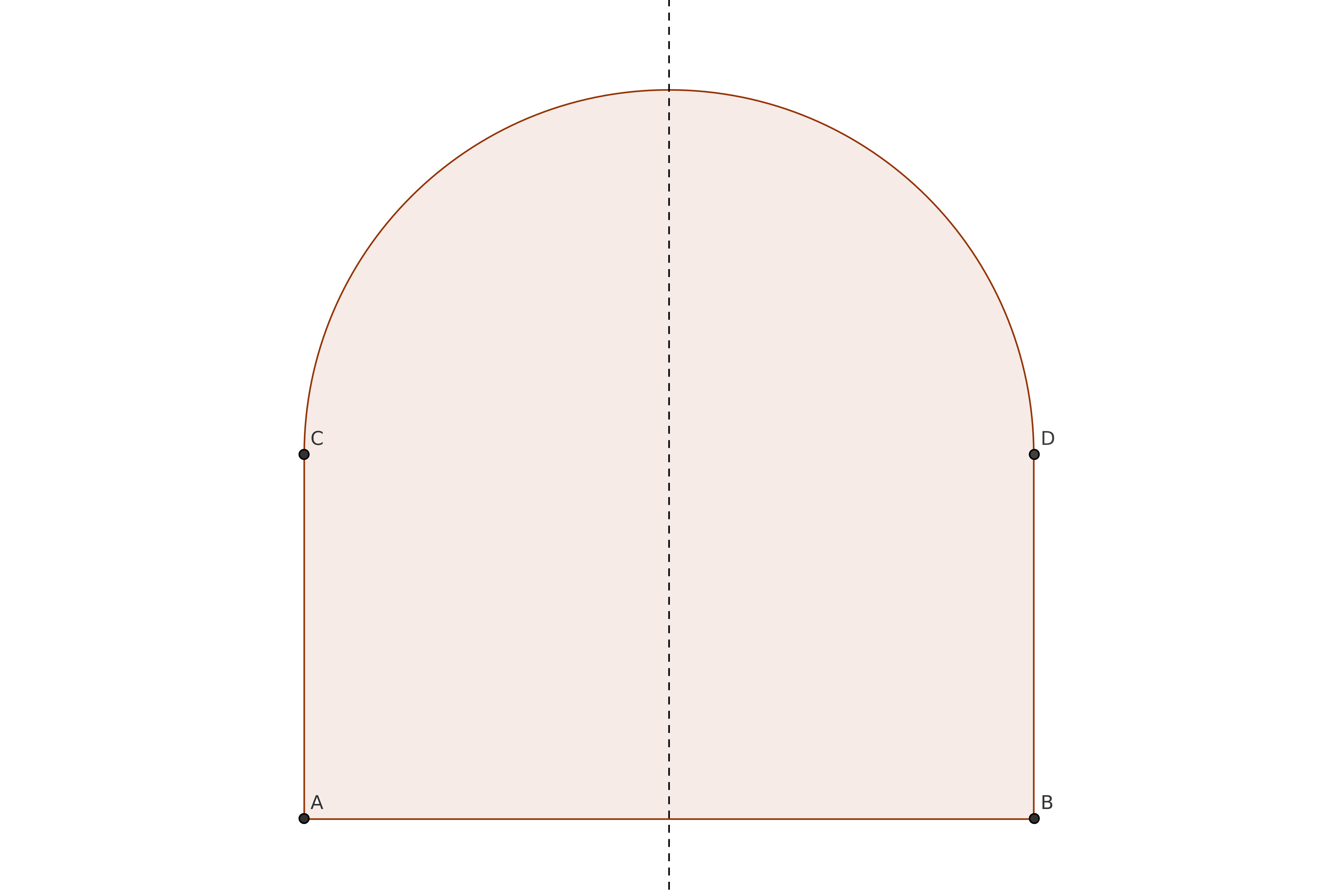}
}

\subfigure[\label{fig:purity} Example of state space leading to a non-canonical theory of purity. In this case, maximal purity is equivalent to maximal resourcefulness: only the pure states are maximally controllable.   However, some pure states are inequivalent resources, meaning that purity is not the only resource into play.      ]{
\includegraphics[scale=0.1]{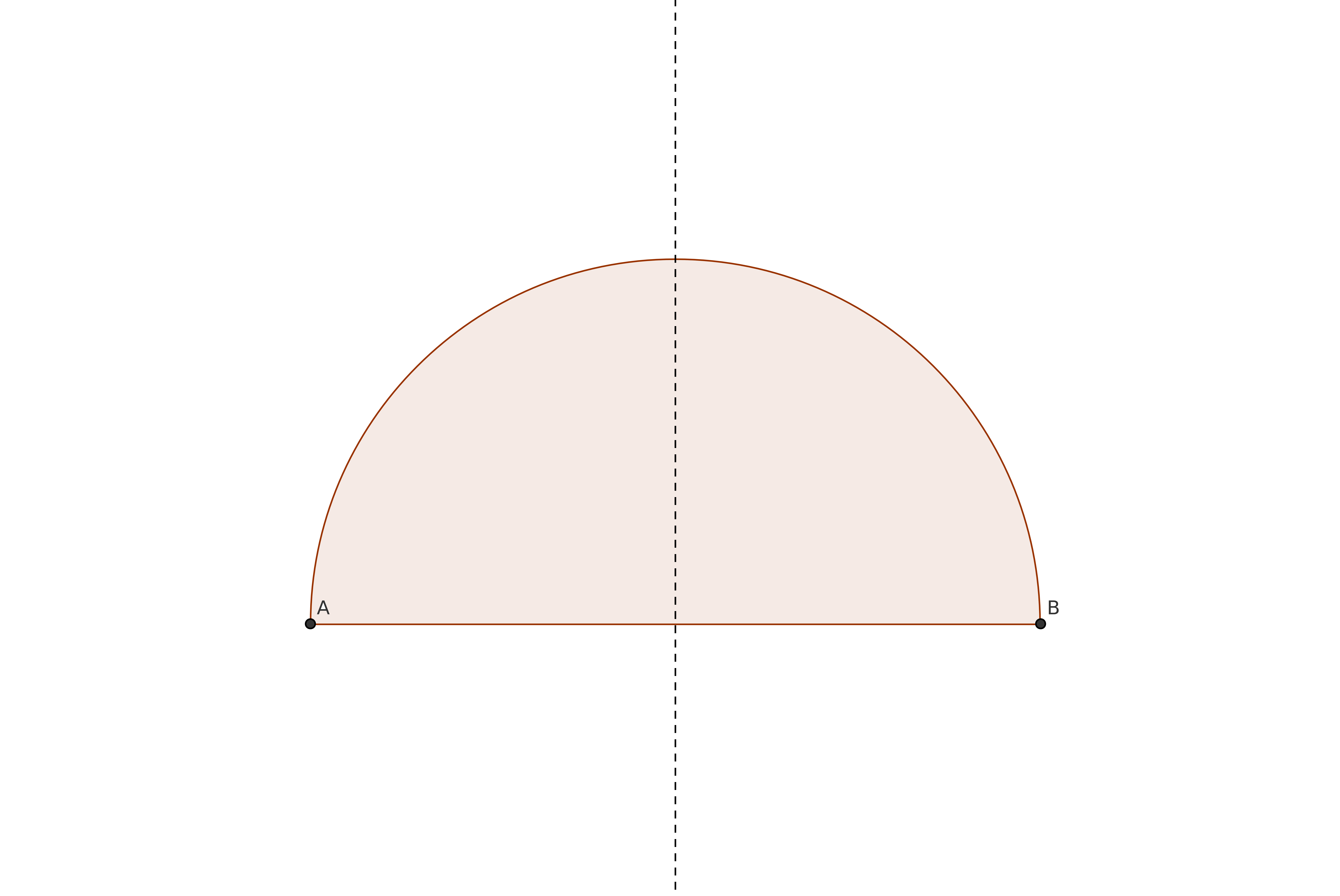}
}

\subfigure[\label{fig:square} Example of state space compatible with a canonical theory
of purity. Here the set of maximally controllable states coincides with the set of pure states, and, in addition, all pure states are equivalent resources.   ]{
\includegraphics[scale=0.1]{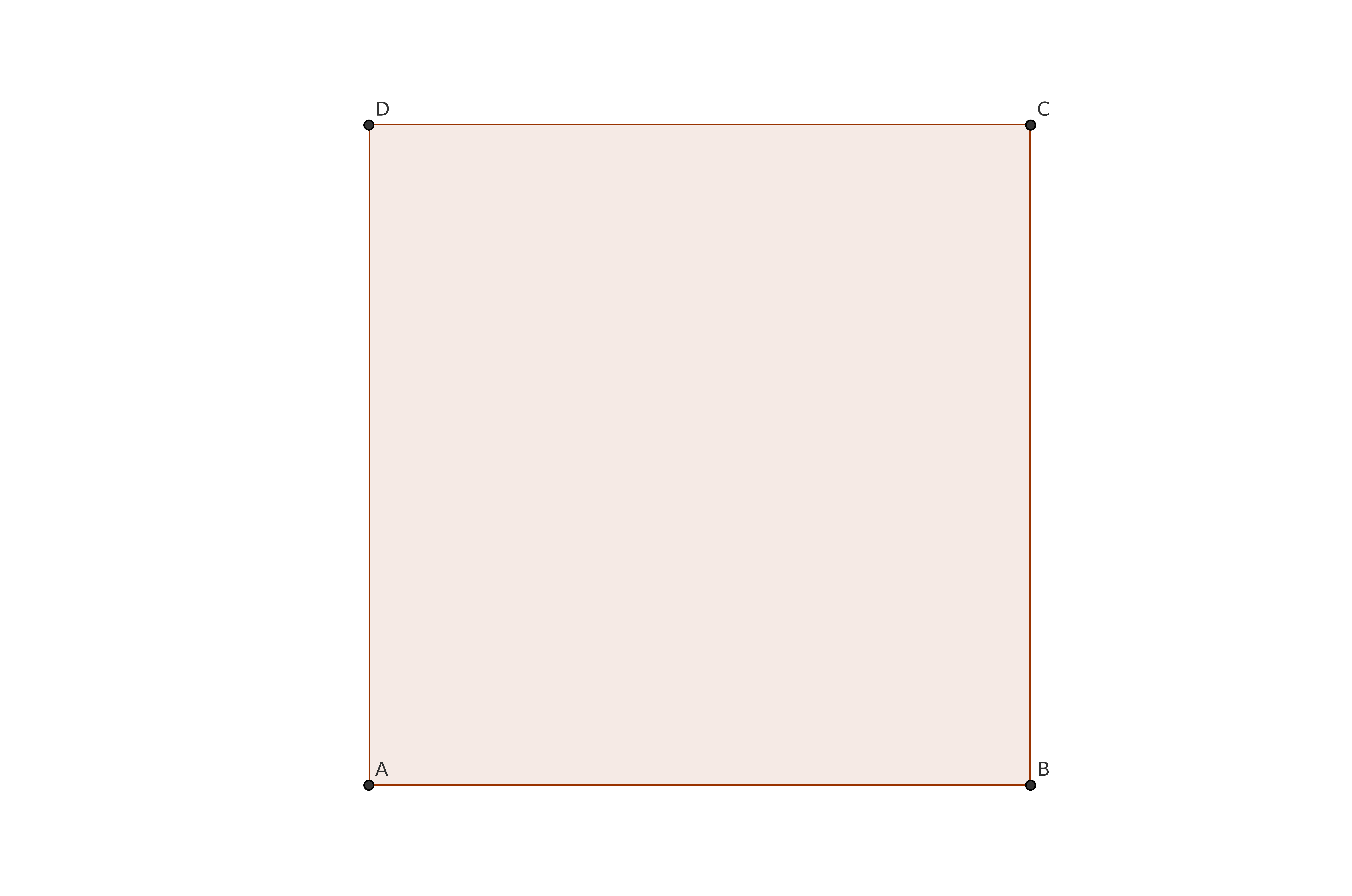}
}

\caption{\label{fig:control}Three different examples of theories of dynamical
control.}
\end{figure}

\begin{example}
Consider a system with the state space depicted in Fig.~\ref{fig:house}.   In this case, there are only two reversible transformations, namely the identity and the reflection around the vertical symmetry axis.  As a consequence, there is no way to obtain the mixed states on the two vertical sides  by applying a RaRe channel to a pure state. These states represent a valuable resource, even though they are not pure.  Since some mixed states are a resource,  the resource theory of dynamical control cannot be viewed as a  resource theory of purity.   

As a second example, consider instead a system whose state space is a half-disk, like in Fig.~\ref{fig:purity}.   Also in this case there are only two reversible transformations (the identity and the reflection around the vertical axis).  However, now every mixed state can be generated from some pure state via a RaRe channel.    The state space can be foliated into horizontal segments generated by pure states under the action of RaRe channels.  As a result, the pure states are the most useful resources and one can interpret the relation $\succeq$ as a way to compare the degree of purity of different states.   Nevertheless,  pure states on different segments are inequivalent resources. In this case there are different, inequivalent classes of pure states: purity is not the only relevant resource into play.

Finally, consider a system with a square state space, like in Fig.~\ref{fig:square} and suppose that all the symmetry transformations in the dihedral group $D_4$ are allowed reversible transformations.   In this case, all the pure states are equivalent under reversible transformations and every mixed state can be obtained by applying a RaRe channel to a fixed pure state.     Here, the resource theory of dynamical control becomes a full-fledged resource theory of purity.  
\end{example}

The above examples show that not every operational theory supports a sensible resource theory of purity.  
Motivated by the examples, we put forward the following definition: 
\begin{defi}\label{def:canonical}
A \emph{theory of purity} is a resource theory of dynamical control  where every state $\rho$ can be compared with at least one pure state.       
The theory is called \emph{canonical} if every pure state is comparable to any other pure state. 
\end{defi}  
In this paper we will focus on canonical theories of purity.
 \begin{prop}\label{prop:canonical}
\label{lem:equivalence transitivity mixedness} The following are equivalent:
\begin{enumerate}
\item the theory is a canonical theory of purity
\item \label{enu:transitivity of pure states}   for every system $\rA$, the group of reversible channels acts transitively on the set of pure states.
\item for every system $\rA$, there exists at  least one state that is more controllable than every state.  
\end{enumerate}
\end{prop}
The proof is provided in Appendix~\ref{app:canonical}.

Combining all the statements of proposition~\ref{lem:equivalence transitivity mixedness}, one can see that in a canonical theory of purity every pure state is more controllable than any other  state.

Starting from Hardy's work \cite{Hardy-informational-1}, the transitivity of the action of  reversible channels on pure states has featured in a number of axiomatizations of quantum theory, either directly as an axiom  \cite{Brukner,masanes,masanes2013existence} or  indirectly as a consequence of  an axiom, as in the case of the Purification axiom \cite{Chiribella-informational}.   
Proposition~\ref{lem:equivalence transitivity mixedness} provides a new motivation for this axiom,  now identified as a necessary and sufficient condition for a well-behaved theory of purity.

In a canonical theory of purity, we say that $\rho$ is \emph{purer} than $\rho'$  if  $\rho \succeq \rho'$ and we adopt the notation $\rho \succeq_{\rm pur} \rho'$.    In this case, we also say that $\rho'$ is \emph{more mixed} than $\rho$, denoted by $\rho'\succeq_{\rm mix}   \rho$.        When $\rho    \succeq_{\rm mix} \rho' $ and $\rho'    \succeq_{\rm mix}  \rho$ we say that $\rho$ and $\rho'$ are \emph{equally mixed}, denoted by $\rho  \simeq_{\rm mix} \rho'$. Clearly, every two states that differ by a reversible channel are equally mixed.  
The converse is true for finite dimensional systems, thanks to Eq.~\eqref{reveq}.

\subsection{\label{sub:Mixedness-relation}  Maximally mixed states}

   We say that a state $\chi  \in  \St \left(\rA\right)$ is \emph{maximally mixed}   if it satisfies the property  
\begin{align*}
 \qquad \forall \rho\in\St \left(\rA\right)      :    \quad \rho  \succeq_{\rm mix}    \chi   \quad \Longrightarrow \quad   \rho  =  \chi   \, . 
\end{align*}
Maximally mixed states can be characterized as the states that are invariant under all reversible channels:  
\begin{prop}
A state $\chi  \in  \St \left(\rA\right)$ is maximally mixed if and only if it is invariant, i.e.\ if and only if $\chi  =   \map U \chi$ for every reversible channel $\map U:\rA\to \rA$.  
\end{prop}

We omit the proof, which is straightforward.      Note that  maximally mixed states do not exist in every theory:  for example,  infinite-dimensional quantum  systems have  no maximally mixed density operator, i.e.\ no trace-class operator that is invariant under the action of the full unitary group.     
For finite-dimensional canonical theories, however, the maximally mixed state exists and  is unique under the standard assumption of compactness of the state space  \cite{Chiribella-purification,masanes}.   In this case, the state $\chi$ is not only a maximal element, but also the \emph{maximum} of the relation $  \succeq_{\rm mix}$, namely,   
\begin{align}\label{maxmix}  \chi  \succeq_{\rm mix}  \rho   \qquad \forall \rho\in\St \left(\rA\right)  \, .
\end{align}
This is in analogy with the quantum case, where the maximally mixed state is  given by the  density matrix $\chi  =  I/d$, where $I$ is the identity operator on the system's Hilbert space and $d$ is the Hilbert space dimension.  Another example of finite-dimensional canonical theory  is provided by the square bit:   

\begin{example}  
Consider a system whose state space is a square, as in Fig.~\ref{fig:square} and pick a generic (mixed) state $\rho$. 
The states that are more mixed than $\rho$ are obtained by applying all possible reversible transformations to $\rho$  (i.e.\ all the elements of the dihedral group $D_4$)  and  taking the convex hull of the orbit.  The set of all states that are more mixed than $\rho$ is an octagon, depicted in blue in Fig.~\ref{fig:Mixed square}. All the vertexes of the octagon are equally mixed. The centre of the square is the maximally state $\chi$, the unique invariant state of the system.

\begin{figure}
\includegraphics[scale=0.1]{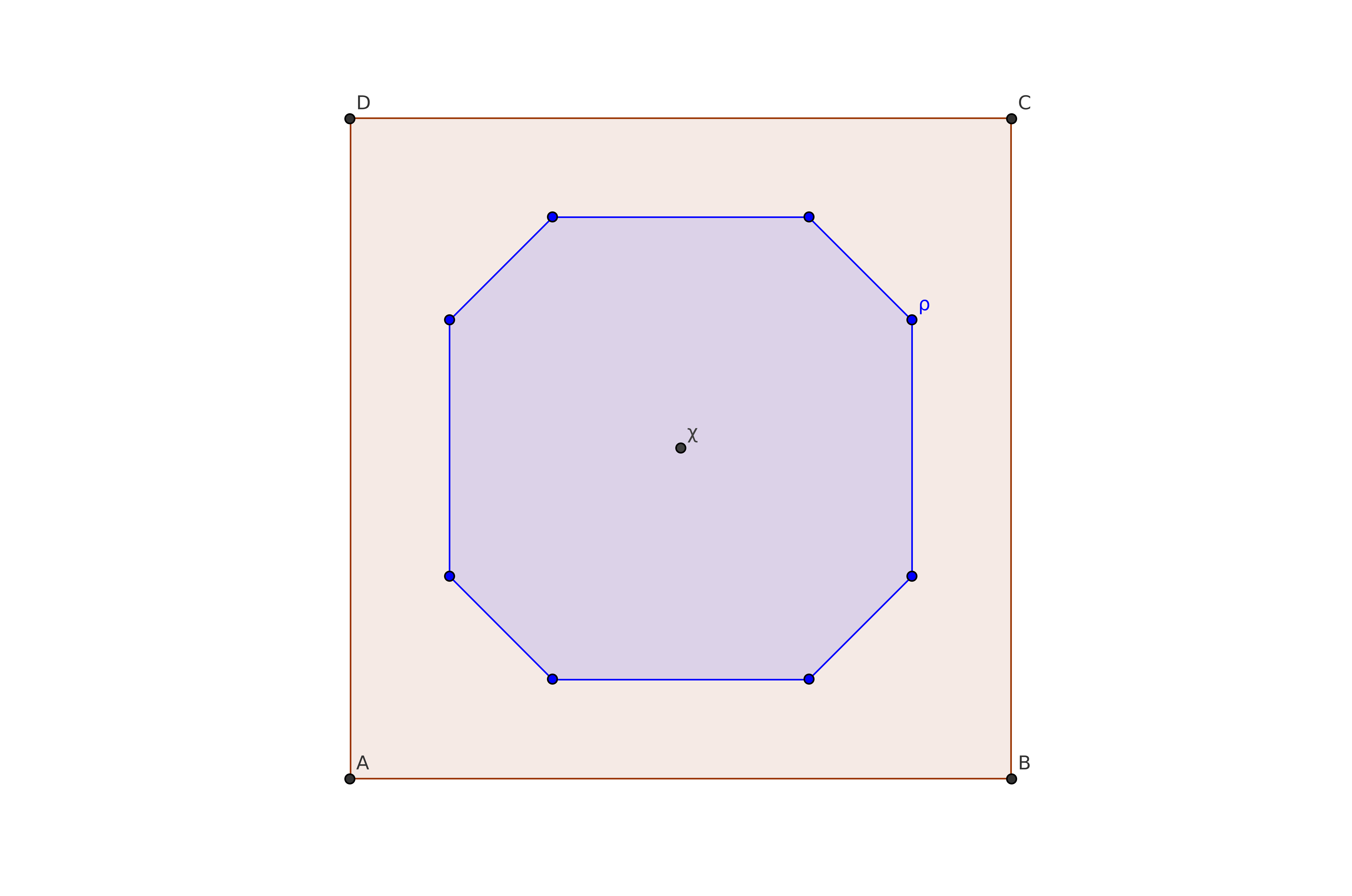}

\protect\caption{\label{fig:Mixed square}Mixedness relation for the state space of a square bit: the vertices of
the octagon represent the states that can be reached from a given
state $\rho$ via reversible transformations. Their
convex hull is the set of states that are more mixed than $\rho$.
Note that it contains the invariant state $\chi$, which can be characterized
as the maximally mixed state.}

\end{figure}

\end{example}

\section{\label{sec:Duality} Entanglement-thermodynamics duality}
In quantum theory, it is well known that the ordering of pure bipartite states according to the degree of entanglement is equivalent to the ordering of their marginals according to the degree of mixedness \cite{Nielsen-Chuang,Nielsen-entanglement,Uhlmann1,Uhlmann2,Uhlmann3}.   In this section we will prove the validity of this equivalence based only on first principles. 

\subsection{Purification} 

In order to establish the desired duality, we consider theories that satisfy the Purification Principle \cite{Chiribella-purification,Chiribella-informational}.      Let us briefly summarize its content.   We say that a state $\rho  \in  \St \left(\rA\right)$ has a \emph{purification}  if there exists a system $\rB$ and a pure state $\Psi\in  \Pur\St \left(\rA\otimes \rB\right)$  (the purification) such that  
\[
\begin{aligned}\Qcircuit @C=1em @R=.7em @!R { & \multiprepareC{1}{\Psi}    & \qw \poloFantasmaCn{\rA} &  \qw    \qquad &  =  & \qquad  &  \prepareC {\rho}  & \qw \poloFantasmaCn{\rA}  &  \qw  \\  & \pureghost{\Psi}    & \qw \poloFantasmaCn{\rB}  &   \measureD{\Tr}   &&&&&&   &  }\end{aligned}
\]  
We say that the purification is {\em essentially unique} if every other purification $\Psi'$ with the same purifying system $ \rB$   satisfies the condition  \begin{align}\label{uniqueness}
\begin{aligned}\Qcircuit @C=1em @R=.7em @!R { & \multiprepareC{1}{\Psi'}    & \qw \poloFantasmaCn{\rA} &  \qw   \\  & \pureghost{\Psi'}    & \qw \poloFantasmaCn{\rB}  &   \qw }\end{aligned}~=\!\!\!\!\begin{aligned}\Qcircuit @C=1em @R=.7em @!R { & \multiprepareC{1}{\Psi}    & \qw \poloFantasmaCn{\rA} &  \qw & \qw & \qw  \\  & \pureghost{\Psi}    & \qw \poloFantasmaCn{\rB}  &   \gate{\cU}& \qw \poloFantasmaCn{\rB}& \qw  }\end{aligned}~,
\end{align}
for some reversible transformation $\map U:\rB\to \rB$.
With these definitions, the Purification Principle can be phrased as 
\begin{ax}[Purification \cite{Chiribella-purification,Chiribella-informational}]
Every state has a purification.  Every purification is essentially unique. 
\end{ax}

Purification has a number  of important consequences. 
First of all, it implies that the group of reversible transformations acts transitively on the set of pure states:
\begin{prop}[Transitivity]
For every system $\rB$ and every  pair of  pure states $\psi,\psi'\in\mathsf{PurSt}\left(\mathrm{B}\right)$
there exists a reversible channel $\mathcal{U}:  \rB\to \rB$ such
that $\psi'=\mathcal{U}\psi$.\end{prop}

The existence of a reversible transformation connecting $\psi$ and $\psi'$ is a consequence of the essential uniqueness of purification [Eq.~\eqref{uniqueness}], in the special case $\rA  =  \rI$  \cite{Chiribella-purification}.   Since all pure states are equivalent under reversible transformations, every theory with purification gives rise to a \emph{canonical} theory of purity, in the sense of definition \ref{def:canonical}. 
One could take this fact as  a further indication that Purification is a good starting point for a well-behaved thermodynamics. 

Another important consequence of Purification is the existence of entanglement:   
\begin{prop}[Existence of entangled states]\label{prop:pure-product}
For every pair of systems $\rA$ and $\rB$,  a pure state of $\rA\otimes \rB$ is entangled if and only if its marginal on system $\rA$ is mixed.  
\end{prop}
\Proof Let us denote the pure bipartite state by $\Psi$.   If $\Psi$ is not entangled, then it must be a product of two pure states, say $\Psi  = \alpha \otimes \beta$.   Clearly, this implies that the marginal on system $\rA$ is pure. 

Conversely,  suppose that  the marginal of $\Psi$ on system $\rA$ is pure and denote it by $\alpha$.  Then, for every pure state  $\beta'\in\Pur\St \left(\rB\right)$,  the product state $\Psi'  =\alpha \otimes \beta'$ is pure, thanks to Purity Preservation.   Now, $\Psi$ and $\Psi'$ are two purifications of $\alpha$. By the essential uniqueness of purification, one must have $\Psi  =  \left(\map I_\rA\otimes \map U_\rB \right)  \Psi'$ for some reversible transformation $\map U_\rB$ acting on system $\rB$.  Hence, we have $\Psi  =   \alpha \otimes  \beta$, with $\beta=  \map U_\rB  \beta'$.   \qed

Finally, Purification implies the \emph{steering property} \cite{Schrodinger,barnum2013ensemble}, stating that every ensemble decomposition of a given state can be generated by a measurement on the purifying system:    
\begin{prop}[Steering]\label{prop:steering}
Let $\rho  $ be a state of system $\rA$ and let $\Psi\in\mathsf{PurSt} \left(\rA\otimes \rB\right)$ be a purification
of $\rho$.   For every ensemble of states $\{\rho_i\}_{i\in\set X}$ such that $\sum_i   \rho_i   =  \rho$, there exists a measurement  $\{b_i\}_{i\in\set X}$ on the purifying system $\mathrm{B}$
 such that the following relation holds\[
\begin{aligned}\Qcircuit @C=1em @R=.7em @!R { & \prepareC{\rho_i}    & \qw \poloFantasmaCn{\rA} &  \qw   }\end{aligned}~=\!\!\!\!\begin{aligned}\Qcircuit @C=1em @R=.7em @!R { & \multiprepareC{1}{\Psi}    & \qw \poloFantasmaCn{\rA} &  \qw   \\  & \pureghost{\Psi}    & \qw \poloFantasmaCn{\rB}  &   \measureD{b_i }}\end{aligned}  \qquad \forall i\in\set X \, .
\]\end{prop}
See theorem 6  of Ref.~\cite{Chiribella-purification} for the proof.   The steering property will turn out to be essential in establishing the duality between entanglement and thermodynamics.

\subsection{One-way  protocols transforming pure states into pure states}

The operational Lo-Popescu theorem guarantees that every LOCC protocol acting on a pure bipartite input state can be simulated by a one-way protocol.   Purification buys us an extra bonus:   not only is the protocol  one-way, but also all the conditional operations are reversible.

\begin{lemma}
\label{lem:LP reversible}   
Let $\Psi$  and $\Psi'$ be  pure states  of $\rA\otimes \rB$.  Under the validity of Purification and Purity Preservation, every one-way protocol transforming $\Psi$ into $\Psi'$ can be simulated  by a one-way protocol where all conditional operations are reversible. 
\end{lemma}

\Proof
Suppose that $\Psi$ can be transformed into $\Psi'$ via a one-way protocol where  Alice performs a test $\left\{\map A_i\right\}_{i\in\set X}$ and Bob performs a channel $\map B^{\left(i\right)}$ conditional on the outcome $i$.   By definition,  we have   
\[
\sum_{i} \!\!\!\! \begin{aligned}\Qcircuit @C=1em @R=.7em @!R { & \multiprepareC{1}{\Psi}    & \qw \poloFantasmaCn{\rA} &  \gate{\cA_{i}}\ar@{-->}[drr] & \qw \poloFantasmaCn{\rA} &\qw &\qw &\qw \\  & \pureghost{\Psi}    & \qw \poloFantasmaCn{\rB}  & \qw &\qw  &  \gate{\cB^{(i)}} & \qw \poloFantasmaCn{\rB}  &\qw}\end{aligned}~=\!\!\!\!\begin{aligned}\Qcircuit @C=1em @R=.7em @!R { & \multiprepareC{1}{\Psi'}    & \qw \poloFantasmaCn{\rA}  & \qw  \\  & \pureghost{\Psi'}    & \qw \poloFantasmaCn{\rB}  &\qw}\end{aligned}~.
\]
Since $\Psi'$ is pure, this implies that there exists a probability distribution $\left\{p_{i}\right\}$ such that 
\begin{align}\label{aaaa}
\begin{aligned}\Qcircuit @C=1em @R=.7em @!R { & \multiprepareC{1}{\Psi}    & \qw \poloFantasmaCn{\rA} &  \gate{\cA_{i}}\ar@{-->}[drr] & \qw \poloFantasmaCn{\rA} &\qw &\qw &\qw \\  & \pureghost{\Psi}    & \qw \poloFantasmaCn{\rB}  & \qw &\qw  &  \gate{\cB^{(i)}} & \qw \poloFantasmaCn{\rB}  &\qw}\end{aligned}~=p_i \!\!\!\!\begin{aligned}\Qcircuit @C=1em @R=.7em @!R { & \multiprepareC{1}{\Psi'}    & \qw \poloFantasmaCn{\rA}  & \qw  \\  & \pureghost{\Psi'}    & \qw \poloFantasmaCn{\rB}  &\qw}\end{aligned} 
\end{align}
for every outcome $i$.  
Now, without loss of generality  each transformation $\map A_i$ can be assumed to be  pure  (if not, one can always decompose it into pure transformations, thanks to the pure decomposition property).   Then, Purity Preservation guarantees that  the normalized state $\Psi_{i}$ defined  by
\begin{equation}\label{eq:psi_i LP lemma}
\begin{aligned}\Qcircuit @C=1em @R=.7em @!R { & \multiprepareC{1}{\Psi_{i}}    & \qw \poloFantasmaCn{\rA} &\qw \\  & \pureghost{\Psi_{i}}    & \qw \poloFantasmaCn{\rB}  &\qw}\end{aligned}~:=p_i^{-1}\!\!\!\!  \begin{aligned}\Qcircuit @C=1em @R=.7em @!R { & \multiprepareC{1}{\Psi}    & \qw \poloFantasmaCn{\rA} &  \gate{\cA_{i}} & \qw \poloFantasmaCn{\rA} &\qw  \\  & \pureghost{\Psi}    & \qw \poloFantasmaCn{\rB}  & \qw &\qw  &\qw}\end{aligned}
\end{equation}
is pure.   With this definition, Eq.~\eqref{aaaa} becomes \[
\begin{aligned}\Qcircuit @C=1em @R=.7em @!R { & \multiprepareC{1}{\Psi_{i}}    & \qw \poloFantasmaCn{\rA} &\qw &\qw &\qw \\  & \pureghost{\Psi_{i}}    & \qw \poloFantasmaCn{\rB}  &  \gate{\cB^{(i)}} & \qw \poloFantasmaCn{\rB}  &\qw}\end{aligned}~=\!\!\!\!\begin{aligned}\Qcircuit @C=1em @R=.7em @!R { & \multiprepareC{1}{\Psi'}    & \qw \poloFantasmaCn{\rA}  & \qw  \\  & \pureghost{\Psi'}    & \qw \poloFantasmaCn{\rB}  &\qw}\end{aligned}  \, . 
\]
Tracing out system $\rB$ on both sides one obtains 
\begin{align*}
\begin{aligned}\Qcircuit @C=1em @R=.7em @!R { & \multiprepareC{1}{\Psi'}    & \qw \poloFantasmaCn{\rA}  & \qw  \\  & \pureghost{\Psi'}    & \qw \poloFantasmaCn{\rB}  &\measureD{\Tr}}\end{aligned} &~=\!\!\!\! \begin{aligned}\Qcircuit @C=1em @R=.7em @!R { & \multiprepareC{1}{\Psi_{i}}    & \qw \poloFantasmaCn{\rA} &\qw &\qw &\qw \\  & \pureghost{\Psi_{i}}    & \qw \poloFantasmaCn{\rB}  &  \gate{\cB^{(i)}} & \qw \poloFantasmaCn{\rB}  &\measureD{\Tr}}\end{aligned} \\ 
&~= \!\!\!\! \begin{aligned}\Qcircuit @C=1em @R=.7em @!R { & \multiprepareC{1}{\Psi_{i}}    & \qw \poloFantasmaCn{\rA} &\qw \\  & \pureghost{\Psi_{i}}    & \qw \poloFantasmaCn{\rB}   &\measureD{\Tr}}\end{aligned}~,
\end{align*}
the second equality coming from the normalization of the channel $\map B^{(i)}$ (cf.\ proposition~\ref{prop:normalization}).  
Hence, the pure states $\Psi_{i} $ and $\Psi'   $
have the same marginal on $\mathrm{A}$. By the essential uniqueness of Purification, they must differ by
a reversible channel $\mathcal{U}^{\left(i\right)}$ on the purifying
system $\mathrm{B}$, namely 
\begin{align}\label{bbbb}
\begin{aligned}\Qcircuit @C=1em @R=.7em @!R { & \multiprepareC{1}{\Psi_{i}}    & \qw \poloFantasmaCn{\rA} &\qw &\qw &\qw \\  & \pureghost{\Psi_{i}}    & \qw \poloFantasmaCn{\rB}   &\gate{\cU^{(i)}} &\qw \poloFantasmaCn{\rB} &\qw}\end{aligned}~=\!\!\!\!\begin{aligned}\Qcircuit @C=1em @R=.7em @!R { & \multiprepareC{1}{\Psi'}    & \qw \poloFantasmaCn{\rA}  & \qw  \\  & \pureghost{\Psi'}    & \qw \poloFantasmaCn{\rB}  &\qw}\end{aligned}
\end{align}  
In conclusion, we obtained  
\begin{align*}
& \begin{aligned}\Qcircuit @C=1em @R=.7em @!R { & \multiprepareC{1}{\Psi}    & \qw \poloFantasmaCn{\rA} &  \gate{\cA_{i}}\ar@{-->}[drr] & \qw \poloFantasmaCn{\rA} &\qw &\qw &\qw \\  & \pureghost{\Psi}    & \qw \poloFantasmaCn{\rB}  & \qw &\qw  &  \gate{\cB^{(i)}} & \qw \poloFantasmaCn{\rB}  &\qw}\end{aligned} ~ =\qquad \qquad \qquad \\
 &\qquad \qquad \qquad  =   p_i  \!\!\!\! 
 \begin{aligned}\Qcircuit @C=1em @R=.7em @!R { & \multiprepareC{1}{\Psi'}    & \qw \poloFantasmaCn{\rA}  & \qw  \\  & \pureghost{\Psi'}    & \qw \poloFantasmaCn{\rB}  &\qw}\end{aligned}    \\
&\qquad \qquad \qquad  =p_i  \!\!\!\! \begin{aligned}\Qcircuit @C=1em @R=.7em @!R { & \multiprepareC{1}{\Psi_i}    & \qw \poloFantasmaCn{\rA}  &\qw &\qw  &\qw \\  & \pureghost{\Psi_i}    & \qw \poloFantasmaCn{\rB}  &  \gate{\cU^{(i)}} & \qw \poloFantasmaCn{\rB}  &\qw}\end{aligned}  \\
&\qquad \qquad \qquad = \!\!\!\!
  \begin{aligned}\Qcircuit @C=1em @R=.7em @!R { & \multiprepareC{1}{\Psi}    & \qw \poloFantasmaCn{\rA} &  \gate{\cA_{i}}\ar@{-->}[drr] & \qw \poloFantasmaCn{\rA} &\qw &\qw &\qw \\  & \pureghost{\Psi}    & \qw \poloFantasmaCn{\rB}  & \qw &\qw  &  \gate{\cU^{(i)}} & \qw \poloFantasmaCn{\rB}  &\qw}\end{aligned}~,
\end{align*}
where we have used Eqs.~\eqref{aaaa},  \eqref{bbbb}, and  \eqref{eq:psi_i LP lemma}. In other words,  the initial protocol can be simulated by a protocol where Alice performs the test $\left\{\map A_i\right\}$ and Bob performs the reversible transformation $\map U^{\left(i\right)}$ conditionally on the outcome $i$.  
\qed

The reduction to one-way protocols with reversible operations is the key  to connect the resource theory of entanglement with the resource theory of purity.  The duality between these two resource theories will be established in the next subsections.

\subsection{The more entangled a pure state, the more mixed its marginals}

We start by proving one direction of the entanglement-thermodynamics duality: if a state is more entangled than another, then the marginals of the former are more mixed than the marginals of the latter: 

\begin{lemma}
\label{lem:more-entangled -> more-mixed}  
Let $\Psi$ and   $\Psi'$ be two pure states of  system $\rA\otimes \rB$  and let $\rho, \rho'$ and $\sigma,  \sigma'$  be their marginals on system $\rA$   and $\rB$, respectively.  Under the validity of Purification, Purity Preservation, and Local Exchangeability,  if $\Psi$ is more entangled than $\Psi'$, then $\rho$   ($\sigma$) is more mixed than $\rho'$  ($\sigma'$). 
\end{lemma}
\Proof
By the operational Lo-Popescu theorem, we know that there exists a one-way protocol transforming $\Psi$  into $\Psi'$.  Moreover, thanks to Purification, the conditional operations in the protocol can be chosen to be reversible (lemma~\ref{lem:LP reversible}).       Let us choose a protocol with classical communication  from Alice to Bob, in which Alice performs the test $\left\{\map A_i\right\}_{i\in\set X}$ and Bob performs the reversible transformation $\map U^{\left(i\right)}$ conditional on the outcome $i$.     
Since $\Psi'$ is pure, we must have
\[
\begin{aligned}\Qcircuit @C=1em @R=.7em @!R { & \multiprepareC{1}{\Psi}    & \qw \poloFantasmaCn{\rA} &  \gate{\cA_{i}}\ar@{-->}[drr] & \qw \poloFantasmaCn{\rA} &\qw &\qw &\qw \\  & \pureghost{\Psi}    & \qw \poloFantasmaCn{\rB}  & \qw &\qw  &  \gate{\cU^{(i)}} & \qw \poloFantasmaCn{\rB}  &\qw}\end{aligned}~=p_{i}\!\!\!\! \begin{aligned}\Qcircuit @C=1em @R=.7em @!R { & \multiprepareC{1}{\Psi'}    & \qw \poloFantasmaCn{\rA} &\qw \\  & \pureghost{\Psi'}    & \qw \poloFantasmaCn{\rB}    &\qw}\end{aligned}  \qquad \forall i\in\set X~,
\]  
where $\left\{p_i\right\}$ is a suitable probability distribution.  
Denoting by $\map V^{\left(i\right)}$ the  inverse of 
$\mathcal{U}^{\left(i\right)}$ and applying it on both sides of the equation, we obtain\[
\begin{aligned}\Qcircuit @C=1em @R=.7em @!R { & \multiprepareC{1}{\Psi}    & \qw \poloFantasmaCn{\rA} &  \gate{\cA_{i}} & \qw \poloFantasmaCn{\rA} &\qw  \\  & \pureghost{\Psi}    & \qw \poloFantasmaCn{\rB}   & \qw & \qw &\qw } \end{aligned} ~ = 
p_{i}\!\!\!\! \begin{aligned}\Qcircuit @C=1em @R=.7em @!R { & \multiprepareC{1}{\Psi'}    & \qw \poloFantasmaCn{\rA} &\qw&\qw &\qw \\  & \pureghost{\Psi'}    & \qw \poloFantasmaCn{\rB}    &\gate{\map V^{(i)}}& \qw \poloFantasmaCn{\rB} &\qw}\end{aligned}~.
\]   
Summing over all outcomes the equality becomes 
\begin{align}\label{cccc} 
\begin{aligned}    
\Qcircuit @C=1em @R=.7em @!R { & \multiprepareC{1}{\Psi}    & \qw \poloFantasmaCn{\rA} &  \gate{\cA} & \qw \poloFantasmaCn{\rA} &\qw  \\  & \pureghost{\Psi}    & \qw \poloFantasmaCn{\rB}   & \qw & \qw &\qw } 
\end{aligned}~ = \!\!\!\!
\begin{aligned}
\Qcircuit @C=1em @R=.7em @!R { & \multiprepareC{1}{\Psi'}    & \qw \poloFantasmaCn{\rA} &\qw &\qw &\qw \\  & \pureghost{\Psi'}    & \qw \poloFantasmaCn{\rB}    &\gate{\map R }& \qw \poloFantasmaCn{\rB} &\qw}
\end{aligned} ~,
\end{align}
with $\mathcal{A}  :=  \sum_i  \cA_i$ and $\map R  :  = \sum_i p_i  \map V^{\left(i\right)}$.  
Finally, we obtain
\begin{align*}
\begin{aligned}
\Qcircuit @C=1em @R=.7em @!R { 
&  \prepareC{\sigma}  & \qw \poloFantasmaCn{\rB}  & \qw   }  
\end{aligned} &~  
= \!\!\!\! \begin{aligned}    
\Qcircuit @C=1em @R=.7em @!R { & \multiprepareC{1}{\Psi}    & \qw \poloFantasmaCn{\rA} &  \measureD{\Tr} \\  & \pureghost{\Psi}    & \qw \poloFantasmaCn{\rB}   & \qw  } \end{aligned}   \\  &\\
&~= \!\!\!\! \begin{aligned}    
\Qcircuit @C=1em @R=.7em @!R { & \multiprepareC{1}{\Psi}    & \qw \poloFantasmaCn{\rA} &  \gate{\cA} & \qw \poloFantasmaCn{\rA} &\measureD{\Tr} \\  & \pureghost{\Psi}    & \qw \poloFantasmaCn{\rB}   & \qw & \qw &\qw } \end{aligned}   \\  &\\
& ~ = \!\!\!\!
\begin{aligned}
\Qcircuit @C=1em @R=.7em @!R { & \multiprepareC{1}{\Psi'}    & \qw \poloFantasmaCn{\rA} &\qw &\qw &\measureD{\Tr} \\  & \pureghost{\Psi'}    & \qw \poloFantasmaCn{\rB}    &\gate{\map R }& \qw \poloFantasmaCn{\rB} &\qw}
\end{aligned}\\  &\\
&~  = \!\!\!\!         
\begin{aligned}
\Qcircuit @C=1em @R=.7em @!R { 
&  \prepareC{\sigma'}  & \qw \poloFantasmaCn{\rB}  & \gate{\map R}  &  \qw \poloFantasmaCn{\rB}    &\qw }    
\end{aligned}   ~ ,  
\end{align*}
where we have used the normalization of channel $\map A$ in the second equality and Eq.~\eqref{cccc} in the third. 
Since $\map R$ is a RaRe channel by construction, we have proved that $\sigma$ is more mixed than $\sigma'$.  The fact  that $\rho$ is more mixed than $\rho'$ can be proved by the same argument, starting from a one-way protocol with classical communication from Bob to Alice and with reversible operations on Alice's side.  \qed

The relation between degree of entanglement of a pure state and degree of mixedness of its marginals holds not only for bipartite states, but also for multipartite states.   Indeed, suppose that $\Psi$ and $\Psi'$ are two pure states of system $\rA_1\otimes \rA_2\otimes \cdots \otimes \rA_N$ and that $\Psi$ is more entangled than $\Psi'$, in the sense that there exists a (multipartite) LOCC protocol converting $\Psi$ into $\Psi'$.     For every subset $\set S  \subset  \left\{1,\dots, N\right\}$ one can define $\rA   :  =   \bigotimes_{n\not \in  \set S}    \rA_n$  and $\rB:  = \bigotimes_{  n\in\set S} \rA_n$ and apply lemma~\ref{lem:more-entangled -> more-mixed}.
   As a result, one obtains that the marginals of $\Psi$ are more mixed than the marginals of $\Psi'$  \emph{on every subsystem}. 
   
\subsection{\label{sub:More-mixed-implies-more-entangled}  The more mixed a state, the more entangled its purification}

We now prove the converse direction of the entanglement-thermodynamics duality:  if a state is more mixed than another, then its purification is more entangled. 
Remarkably, the proof of this fact requires only the validity of Purification.   
\begin{lemma}\label{lem:more mixed -> more entangled}   
Let $\rho$
and $\rho' $ be two  states of system $\rA$
and let $ \Psi$  ($\Psi' $) be a purification of $\rho$  ($\rho'$), with purifying system $\rB$. 
Under the validity of Purification,
if $\rho$
is more mixed than $\rho'$, then $\Psi$
is more entangled than $\Psi'$.\end{lemma}
\Proof
By hypothesis, one has   
$$    \begin{aligned}\Qcircuit @C=1em @R=.7em @!R { & \prepareC{\rho'}    & \qw \poloFantasmaCn{\rA} &  \gate{\map R}  &\qw \poloFantasmaCn{\rA} &\qw }\end{aligned} ~ = \!\!\!\!  \begin{aligned}\Qcircuit @C=1em @R=.7em @!R { & \prepareC{\rho}    & \qw \poloFantasmaCn{\rA} &  \qw   }\end{aligned} $$ for some RaRe channel $\map R:  =  \sum_{i}p_{i}  \, \mathcal{U}^{\left(i\right)}$. 
Let us define the bipartite state $\Theta$ as
\begin{align}
\nonumber
\begin{aligned}\Qcircuit @C=1em @R=.7em @!R { & \multiprepareC{1}{\Theta}    & \qw \poloFantasmaCn{\rA} &\qw \\  & \pureghost{\Theta}    & \qw \poloFantasmaCn{\rB}  & \qw } \end{aligned} & ~:=\!\!\!\! \begin{aligned}\Qcircuit @C=1em @R=.7em @!R { & \multiprepareC{1}{\Psi'}    & \qw \poloFantasmaCn{\rA} &\gate{\map R} & \qw \poloFantasmaCn{\rA} &\qw \\  
& \pureghost{\Psi'}    & \qw \poloFantasmaCn{\rB}  & \qw &\qw &\qw } \end{aligned}\\
\nonumber &\\
\label{theta}
&  ~\equiv  
 \sum_i  p_i\!\!\!\! \begin{aligned}\Qcircuit @C=1em @R=.7em @!R { & \multiprepareC{1}{\Psi'}    & \qw \poloFantasmaCn{\rA} &\gate{\map U^{(i)}} & \qw \poloFantasmaCn{\rA} &\qw \\  & \pureghost{\Psi'}    & \qw \poloFantasmaCn{\rB}  & \qw &\qw &\qw } \end{aligned} 
\end{align}
By construction, $\Theta$ is an extension of $\rho$: indeed, one has
\begin{align*}
\begin{aligned}\Qcircuit @C=1em @R=.7em @!R { & \multiprepareC{1}{\Theta}    & \qw \poloFantasmaCn{\rA} &\qw \\  & \pureghost{\Theta}    & \qw \poloFantasmaCn{\rB}  &\measureD{\Tr} } \end{aligned} 
&~=
\!\!\!\! \begin{aligned}\Qcircuit @C=1em @R=.7em @!R { & \multiprepareC{1}{\Psi'}    & \qw \poloFantasmaCn{\rA} &\gate{\map R} & \qw \poloFantasmaCn{\rA} &\qw \\  & \pureghost{\Psi'}    & \qw \poloFantasmaCn{\rB}  &\qw &\qw & \measureD{\Tr} } \end{aligned} \\ 
&\\
&~=\!\!\!\! \begin{aligned}\Qcircuit @C=1em @R=.7em @!R { & \prepareC{\rho'}    & \qw \poloFantasmaCn{\rA} &  \gate{\map R}  &\qw \poloFantasmaCn{\rA} &\qw }\end{aligned}\\
&\\
& ~ =\!\!\!\! \begin{aligned}\Qcircuit @C=1em @R=.7em @!R { & \prepareC{\rho}    & \qw \poloFantasmaCn{\rA} &  \qw   }\end{aligned}~.
\end{align*}
Let us take a purification of $\Theta$, say   $\Gamma\in\Pur\St \left(\rA\otimes\rB\otimes \rC\right)$.  Clearly, $\Gamma$ is a purification of $\rho$, since one has
\begin{align*}
\begin{aligned}
\Qcircuit @C=1em @R=.7em @!R { 
& \multiprepareC{2}{\Gamma}    & \qw \poloFantasmaCn{\rA} &\qw 
\\  & \pureghost{\Gamma}    & \qw \poloFantasmaCn{\rB}  &\measureD{\Tr} \\ & \pureghost{\Gamma}    & \qw \poloFantasmaCn{\rC}  &\measureD{\Tr}} \end{aligned}
 ~ =\!\!\!\!
\begin{aligned}\Qcircuit @C=1em @R=.7em @!R { & \multiprepareC{1}{\Theta}    & \qw \poloFantasmaCn{\rA} &\qw \\  & \pureghost{\Theta}    & \qw \poloFantasmaCn{\rB}  &\measureD{\Tr} } \end{aligned}   
~  = \!\!\!\!
  \begin{aligned}\Qcircuit @C=1em @R=.7em @!R { & \prepareC{\rho}    & \qw \poloFantasmaCn{\rA} &  \qw   }\end{aligned}~. 
\end{align*}
Then, the essential uniqueness of purification implies that $\Gamma$ must be of the form  
\begin{align}\label{gamma}
\begin{aligned}\Qcircuit @C=1em @R=.7em @!R { & \multiprepareC{2}{\Gamma}    & \qw \poloFantasmaCn{\rA} &\qw \\  & \pureghost{\Gamma}    & \qw \poloFantasmaCn{\rB}  &\qw \\ & \pureghost{\Gamma}    & \qw \poloFantasmaCn{\rC}  &\qw} \end{aligned} ~=\!\!\!\! \begin{aligned}\Qcircuit @C=1em @R=.7em @!R { & \multiprepareC{1}{\Psi}    & \qw \poloFantasmaCn{\rA} &\qw &\qw &\qw \\  & \pureghost{\Psi}    & \qw \poloFantasmaCn{\rB}  &\multigate{1}{\cU} &\qw \poloFantasmaCn{\rB} &\qw \\ &\prepareC{\gamma} & \qw \poloFantasmaCn{\rC} &\ghost{\cU} &\qw \poloFantasmaCn{\rC} &\qw} \end{aligned}~,
\end{align}
for some reversible transformation $\map U$  and some pure state $ \gamma$. 
In other words,  $\Psi$ can be transformed into $\Gamma$ by local operations on Bob's side. 

Now, Eq.~\eqref{theta}  implies that  the states $  \left\{  p_i     \left(\map U^{\left(i\right)}  \otimes \map I_\rB \right) \Psi  \right\}_{i\in\set X}$  are an ensemble decomposition of $\Theta$.  
Hence, the steering property (proposition~\ref{prop:steering}) implies that there exists a measurement $\left\{c_i\right\}_{i\in\set X}$ on $ \rC $ such that
\begin{align}\label{steer}
p_{i}\!\!\!\! \begin{aligned}\Qcircuit @C=1em @R=.7em @!R { & \multiprepareC{1}{\Psi'}    & \qw \poloFantasmaCn{\rA} &\gate{\cU_{i}} & \qw \poloFantasmaCn{\rA} &\qw \\  & \pureghost{\Psi'}    & \qw \poloFantasmaCn{\rB}  & \qw &\qw &\qw } \end{aligned}  
~ =  \!\!\!\! 
\begin{aligned}\Qcircuit @C=1em @R=.7em @!R { & \multiprepareC{2}{\Gamma}    & \qw \poloFantasmaCn{\rA} &\qw \\  & \pureghost{\Gamma}    & \qw \poloFantasmaCn{\rB}  &\qw \\ & \pureghost{\Gamma}    & \qw \poloFantasmaCn{\rC}  &\measureD{c_{i}}} \end{aligned}   \qquad \forall i\in\set X  \,.
\end{align}
Combining Eqs.~\eqref{gamma} and \eqref{steer}, we obtain the desired result.
\begin{align*} \begin{aligned}\Qcircuit @C=1em @R=.7em @!R{ & \multiprepareC{1}{\Psi'}    & \qw \poloFantasmaCn{\rA} & \qw \\  & \pureghost{\Psi'}    & \qw \poloFantasmaCn{\rB}  & \qw  } \end{aligned}  
&  =  \sum_{i\in\set X}   \!\!\!\! 
 \begin{aligned} 
\Qcircuit @C=1em @R=.7em @!R { & \multiprepareC{1}{\Psi}    & \qw \poloFantasmaCn{\rA}  & \qw  &\qw &\gate{{\cU}^{(i)}} &\qw \poloFantasmaCn{\rA'}  & \qw \\  & \pureghost{\Psi}    & \qw \poloFantasmaCn{\rB}  & \gate{{\cB}_i}\ar@{-->}[urr] &\qw  \poloFantasmaCn{\rB'}  &\qw &\qw &\qw}\end{aligned}     ~,
\end{align*}  
where $\left\{\map B_i\right\}_{i\in\set X}$ is the test defined by  
$$    
\begin{aligned}
\Qcircuit @C=1em @R=.7em @!R {&\qw \poloFantasmaCn{\rB}  & \gate{{\cB}_i}  &\qw  \poloFantasmaCn{\rB}  &\qw} 
\end{aligned}  ~ : = \!\!\!\!
\begin{aligned}
\Qcircuit @C=1em @R=.7em @!R {&  & \qw \poloFantasmaCn{\rB}  &\multigate{1}{\cU} &\qw \poloFantasmaCn{\rB} &\qw\\ 
&\prepareC{\gamma} & \qw \poloFantasmaCn{\rC} &\ghost{\cU} &\qw \poloFantasmaCn{\rC} &\measureD{c_i}} 
\end{aligned}
 $$
In conclusion, if the marginal state  of $\Psi$   is more mixed than the marginal state of $\Psi'$, then $\Psi$ can be converted into $\Psi'$ by a one-way LOCC protocol.  \qed

\subsection{The duality}

Combining lemmas~\ref{lem:more-entangled -> more-mixed}  and    \ref{lem:more mixed -> more entangled}  we identify the degree of entanglement of a pure bipartite state with the degree of mixedness of its marginals:    
\begin{theo}\label{theo:duality}
{\bf (Entanglement-thermodynamics duality)}
Let $\Psi $ and $\Psi $
be two  pure states of system $\rA\otimes \rB$ and let $\rho$, $\rho'$ and $\sigma$, $\sigma'$ be their marginals on system $\rA$ and $\rB$, respectively. 
Under the validity of Purification, Purity Preservation, and Local Exchangeability,  the following statements are equivalent:
\begin{enumerate}
\item $\Psi$ is more entangled than $\Psi'$
\item $\rho$ is more mixed than $\rho'$
\item $\sigma$ is more mixed than $\sigma'$.
\end{enumerate}
\end{theo}
\Proof  The implications $1\Longrightarrow  2$ and  $1\Longrightarrow  3$ follow from lemma~ \ref{lem:more-entangled -> more-mixed} and require the validity of all the three axioms.   
 The implications   $2\Longrightarrow  1$ and  $3\Longrightarrow  1$ follow from lemma~\ref{lem:more mixed -> more entangled}   and require only the validity of Purification.   \qed

The duality can be  illustrated by the commutative diagrams
\[
\xymatrix{
\Psi   \ar[d]_{\mathrm{Tr}_{\rB}} \ar[r]^{\textrm{LOCC}}& \Psi' \ar[d]^{{\Tr_{\rB}}} \\
\rho & \rho'  \ar[l]_{\textrm{RaRe}}}  
\qquad 
\xymatrix{
\Psi   \ar[d]_{\mathrm{Tr}_{\rA}} \ar[r]^{\textrm{LOCC}}& \Psi'\ar[d]^{{\Tr_{\rA}}} \\
\sigma & \sigma' \ar[l]_{\textrm{RaRe}}}    
\]

\noindent  and  is implemented operationally by discarding one of the component systems.     Another illustration of the duality is via the diagram 
\[ \xymatrix{
\Psi     & \Psi' \ar[l]_{\textrm{LOCC}}   \\
  \ar@{-->}[u]^{\mathrm{purification}}   \rho     \ar[r]^{\textrm{RaRe}} & \rho'   \ar@{-->}[u]_{{\rm purification}} }  \, .
\]
Here the map implementing the duality is (a choice of)  purification.  Such a map cannot be realized as a physical operation  \cite{Chiribella-purification}. Instead, it corresponds  to the theoretical operation  of modelling mixed states as marginals of pure states.   

\section{\label{sec:consequences}Consequences of the duality}
In this section we  discuss the simplest consequences of the entanglement-thermodynamics duality, including the relation between maximally mixed and  maximally entangled states, as well as a link between information erasure and generation of entanglement.      
From now on,  the axioms used to derive the duality   will be treated as standing assumptions and will not be written explicitly in the statement of the results.

\subsection{Equivalence under local reversible transformations}  

Let us start from the easiest consequence of the duality:
\begin{cor}\label{cor:LU}
Let  $\Psi$ and $\Psi'$ be two states of system  $\rA\otimes \rB$, with  $\rA$  finite-dimensional.  Then,  $\Psi$ and $\Psi'$  are equally entangled if and only if they are equivalent under local reversible transformations, namely
 \[  \Psi'  =  \left(\map U\otimes \map V\right)   \Psi  \]  
 where    $\map U$ and $\map V$ are reversible transformations acting on $\rA$ and $\rB$, respectively.   
\end{cor}
This result, proved  in appendix~\ref{app:LU},    guarantees that the equivalence classes under the entanglement relation have a simple structure, inherited from the reversible dynamics allowed by the theory.   For finite systems, pure bipartite entanglement is completely characterized by the quotient of the set of pure states under local reversible transformations.

\subsection{Duality for states on  different systems}
Theorem~\ref{theo:duality} concerns the convertibility of  states of the same system. 
To generalize it to arbitrary systems, it is enough to observe that the tensor product with local pure states does not change the degree of entanglement:   for arbitrary  pure states $\Psi$,  $\alpha'$, and $\beta'$ of systems $\rA\otimes \rB$, $\rA'$, and $\rB'$ one has 
\begin{align}\label{equaent}
\Psi      \simeq_{\rm ent}         \alpha'  \otimes \Psi  \otimes \beta'\, ,
\end{align}  
relative to the bipartition  $\left(\rA'\otimes \rA\right)\otimes \left(\rB\otimes \rB'\right)$.  As a consequence, one has the equivalence
 \begin{align*}
\Psi  \succeq_{\rm ent}  \Psi' \quad \Longleftrightarrow  \quad    \alpha'  \otimes \Psi  \otimes \beta'      \succeq_{\rm ent}   \alpha  \otimes \Psi'  \otimes \beta    
\end{align*} 
for arbitrary pure states $\alpha,  \alpha',\beta,\beta'$ of $\rA,\rA' ,\rB,\rB'$, respectively.  
This fact leads directly to the generalization of the duality to states of different systems: 
\begin{cor}
Let $\Psi  $ and $\Psi' $
be two pure states of systems $\rA\otimes \rB$ and  $\rA'\otimes\rB'$, respectively, and  let $\rho$, $\rho'$, $\sigma$ and $\sigma'$ be their marginals on system $\rA,\rA',\rB$ and $\rB'$ respectively. 
Under the validity of Purification, Purity Preservation, and Local Exchangeability,  the following statements are equivalent:
\begin{enumerate}
\item $\Psi$ is more entangled than $\Psi'$
\item $\rho  \otimes \alpha'$ is more mixed than $\alpha  \otimes \rho' $ for every pair of pure states $\alpha \in\Pur\St \left(\rA\right)$ and $\alpha'\in\Pur\St \left(\rA'\right)$.
\item $\sigma\otimes \beta'$ is more mixed than $\beta\otimes \sigma'$ for  every pair of pure states $\beta \in\Pur\St \left(\rB\right)$ and $\beta'\in\Pur\St \left(\rB'\right)$.
\end{enumerate}
\end{cor}

The duality is now implemented by the operation of discarding systems and preparing pure states, as   illustrated by the commutative diagrams
\[
\xymatrix{
\Psi   \ar[d]_{   \mathrm{Tr}_{\rB}  \otimes \alpha'} \ar[r]^{\textrm{LOCC}}& \Psi' \ar[d]^{{  \alpha\otimes \Tr_{\rB}}} \\
\rho  \otimes \alpha' & \alpha\otimes \rho'  \ar[l]_{\textrm{RaRe}}}  
\qquad 
\xymatrix{
\Psi   \ar[d]_{\mathrm{Tr}_{\rA} \otimes \beta'} \ar[r]^{\textrm{LOCC}}& \Psi'\ar[d]^{{\beta\otimes \Tr_{\rA} }} \\
\sigma \otimes \beta'&  \beta\otimes\sigma' \ar[l]_{\textrm{RaRe}}}    
\]

At this point, a cautionary remark is in order.    Inspired by Eq.~\eqref{equaent} one may be tempted compare the degree of mixedness of states of different systems, by  postulating the relation   
\begin{align}\label{trivial}  \rho   \simeq_{\rm mix}    \rho\otimes \alpha'   
\end{align}
for arbitrary states $\rho$ and arbitrary pure states $\alpha'$. 
The appeal of this choice is that  the duality would maintain  the simple form 
\[
\Psi   \succeq_{\rm ent}  \Psi'      \quad \Longleftrightarrow  \quad    \rho   \succeq_{\rm mix}   \rho'   ,    
\]
even for  states of different  systems.  However, Eq.~\eqref{trivial}  would trivialize the resource theory of purity: as a special case, it would imply the relation $ 1 \simeq_{mix}   \alpha' $ for a generic pure state   $\alpha'$,  meaning that pure states can be freely generated.    Since in a canonical theory of purity  pure states are the most resourceful, having pure states for free would mean having every state for free.  

Another way to compare states of different systems according to their degree of purity would be to postulate the relation
\begin{align}\label{invfree}   \rho   \simeq_{\rm pur}   \rho  \otimes \chi_\rB  , 
\end{align}
where $\chi_\rB$ is the maximally mixed state of system $\rB$ (assuming that such a state exists). 
The rationale for this choice would be that $\chi_\rB$ is the ``minimum-resource state'' in the resource  theory of purity and therefore one may want to consider it  as free.    
This choice would not trivialize the resource theory of purity, but would  break the duality with the resource theory of entanglement.   Indeed, Eq.~\eqref{invfree} would imply as a special case $   1  \simeq_{\rm pur}   \chi_\rB$, meaning that maximally mixed states can be freely generated from nothing.  Clearly, this is not the case for their purifications, which are entangled and cannot be generated freely by LOCC.     
In summary, refraining from comparing mixed states on different systems seems to be the best way  to approach the  duality between the resource theory of entanglement and the resource theory of purity.

\subsection{Measures of mixedness and measures of entanglement}

The duality provides the foundation for the definition of quantitative measures of entanglement.  
In every resource theory, one can define measures of ``resourcefulness'', by introducing functions that are non-increasing under the set of free operations \cite{Spekkens}.  In the resource theory of  entanglement,  this leads to the notion of  entanglement monotones: 
\begin{defi}     An \emph{entanglement monotone} for system $\rA\otimes \rB$ is  a function  $E:  \St \left(\rA\otimes \rB\right)  \to \R$ satisfying the condition  
\[
E\left(\rho\right)   \ge E\left(\rho'\right)   \qquad \forall  \rho,  \rho'  \in \St \left(\rA\otimes \rB\right) \, ,  \rho \succeq_{\rm ent}   \rho'   \, .
\]
\end{defi}
More generally, one may want to compare entangled states on different systems. In this case,  an entanglement monotone  $E$ is a \emph{family} of  functions $E=  \left\{E_{\rA\otimes \rB}~|~  \rA,  \rB  \in  \Sys\right\}$ satisfying the condition 
\[ 
E_{\rA\otimes \rB}   \left(\rho\right)   \ge  E_{\rA'\otimes \rB'}   \left(\rho'\right)
\] for every pair of states $\rho\in\St \left(\rA\otimes \rB\right)$ and $\rho' \in  \St \left(\rA'\otimes \rB'\right)$ satisfying $\rho  \succeq_{\rm ent}  \rho'$.

Similarly, one can define monotones in the resource theory of purity: 
\begin{defi} A \emph{purity monotone} for system $\rA$ is a function $P:  \St \left(\rA\right)  \to \R$ satisfying the condition 
\[
P\left(\rho\right)   \ge P\left(\rho'\right)   \qquad \forall  \rho,  \rho'  \in \St \left(\rA\right) \, ,  \rho \succeq_{\rm pur}   \rho'   \, .
\]
\end{defi}
Recall that in our resource theory of purity we abstain from comparing states on different systems, for the reasons discussed in the end of the previous subsection.   
 Purity monotones give a further indication that the definition of purity in terms of random reversible channels is a sensible one:  indeed, if we restrict our attention to the classical case,  the notion of purity monotone introduced here coincides with the canonical notion of Schur-convex function in the theory of majorization \cite{olkin} (see appendix~\ref{app:schur}).  Schur-convex functions are the key tool to construct entropies and other measures of mixedness in classical statistical mechanics, and have  applications in a number of diverse fields \cite{arnold2007}.        

Constructing purity monotones is fairly easy.   For example, every function that is convex and invariant under reversible transformations is a purity monotone:   
\begin{prop}\label{prop:monotones}
Let $P:\St \left(\rA\right) \to \R$ be a function satisfying 
\begin{enumerate}
\item \emph{convexity:}   $ P \left( \sum_i    p_i   \rho_i  \right)   \le \sum_i     p_i     P\left(\rho_i \right)$
for every set of states $\left\{\rho_i\right\}$ and for every probability distribution $\left\{p_i\right\}$, and  
\item \emph{invariance under reversible transformations:}   $P\left(\map U  \rho\right)   =  P\left(\rho\right)$ for every state $\rho$ and for every reversible transformation $\map U$.
\end{enumerate}
Then, $P$ is a purity monotone. 
\end{prop}
The proof is elementary and is  presented in the appendix \ref{app:schur}  for the convenience of the reader.   We highlight that the above proposition is  the natural extension of a well-known result in majorization theory, namely that every convex function that is symmetric in its variables is automatically Schur-convex \cite{olkin}.  Again, it is worth highlighting the perfect match of the operational notions discussed here with the canonical results about majorization. 

   Using proposition~\ref{prop:monotones}, one can construct purity monotones aplenty:     for every convex function $f:  \R \to \R$ one can define the \emph{$f$-purity}   $P_f :  \St \left(\rA\right)  \to \R$ as  
\[
P_f  \left(\rho\right)    :=  \sup_{  \bs a  \,  {\rm pure}    }    \sum_{x\in\set X}    f   \left(  p_x  \right)  \qquad p_x  : =  \left(a_x|\rho\right) \, ,
\]
where the supremum runs over all pure measurements  $\bs a  =  \left\{a_x\right\}_{x\in\set X}$ and over all outcome spaces $\set X$.   It is easy to verify that every $f$-purity is convex and invariant under reversible transformations, and therefore is a purity monotone.   In the special case of the function $f\left(x\right)  =   x\log x $, one has 
\begin{align}\label{measent}  P_f\left(\rho\right)  =   -H  \left(\rho\right) \, ,
\end{align} where $H$ is the \emph{measurement entropy} \cite{Entropy-Barnum,Entropy-Short,Entropy-Kimura}, namely the minimum over all pure measurements of the Shannon entropy of the  probability distribution resulting from the measurement.  
In the case of $f\left(x\right)  =  x^2$ one obtains instead an generalized notion of ``purity'',  which in the quantum case coincides with the usual notion $P\left(\rho\right)  =  \Tr\left ( \rho^2\right)$.

Another way to construct purity monotones is by using norms on the state space: thanks to proposition \ref{prop:monotones}, every  norm that is invariant under reversible transformations leads to a purity monotone.  
For systems that have an invariant state, an easy example is given by the \emph{operational distance}
\[
P \left(\rho\right) :=  \frac {1}{2}    \left\Vert \rho   -  \chi \right\Vert       \, ,    
\]
where $\left\Vert  \cdot \right\Vert$ is the \emph{operational norm}, defined as  $\left\Vert  \delta  \right\Vert    :  =    \sup_{a_0\in\Eff  \left(\rA\right) }  \left (a|\delta\right)  -   \inf_{a_1\in\Eff \left(\rA \right)}    \left(a_1|\delta\right)$  \cite{Chiribella-purification}, and $\chi$ is the invariant state.    
Another example of purity monotone induced by a norm   is  the notion of purity introduced in Refs.~\cite{Dahlsten,Muller-blackhole},  based on the Schatten 2-norm.  In the quantum case, this notion of purity coincides with the ordinary notion $P\left(\rho\right)  =  \Tr\left ( \rho^2\right)$ and therefore coincides with the $f$-purity with $f\left(x\right)  =  x^2$.   It is not a priori clear whether the 2-norm purity coincides with the $x^2$-purity for more general theories.  

Now,  thanks to the duality we can turn   every   purity monotone into an entanglement monotone.  Given a purity monotone $P:\St \left(\rA\right) \to \R$, we can define 
 the pure state entanglement monotone $E:  \Pur\St \left(\rA\otimes \rB\right)  \to \R$ as \begin{align}\label{puremonotone}
E \left(\Psi\right)    :=     g   \left[  P\left(\rho\right)   \right]  \, ,  \qquad \rho  =  \Tr_\rB   \Psi \, ,      
\end{align}
where $g:  \R \to \R$ is any monotonically decreasing function ($f\left(x\right)  \le  f\left(y\right)$ for $x > y$).  Here the monotonically decreasing behaviour of $g$  implements  the reversing of arrows in the duality.   Furthermore, if the functions $P$ and $f$ have suitable convexity properties, the entanglement monotone can be extended from pure states to arbitrary states using the \emph{convex roof construction}  \cite{LOCC2,Vidal,Plenio}.      
Specifically,  one has the following 
\begin{cor}
 Let $P:\St\left(\rA\right) \to \R$ be a convex purity monotone,  $g:  \R\to \R$ be  a concave,  monotonically decreasing function, and $ E  :  \Pur\St \left(\rA\otimes \rB\right) \to \R$ be the pure state entanglement monotone defined in Eq.~\eqref{puremonotone}.     Then,  the convex roof extension   $E:  \St \left(\rA\otimes \rB\right)  \to \R$ defined by 
\begin{align*}
E \left(\Sigma\right) :  =  \inf_{ \begin{array}{c}  \scriptstyle{\left\{p_i   ,  \Psi_i  \right\}  } \\     \scriptstyle{\sum_i p_i  \Psi_i  =  \Sigma }  
\end{array}}  
\,    \sum_i p_i         E \left( \Psi_i \right) 
\end{align*}
is a convex entanglement monotone.  
\end{cor}  
The proof is the same as in the quantum case \cite{Vidal}.  An easy way to generate entanglement measures is to pick an $f$-purity and  take its negative, which corresponds to the choice $g\left(x\right)  =  - x$.   For example, the choice $f\left(x\right)  =  x\log x$  leads to a generalization  of the  \emph{entanglement of formation} \cite{LOCC2} to all theories satisfying the duality. 

\subsection{Maximally entangled states}

As a consequence of the duality,  there exists a correspondence between maximally mixed  and ``maximally entangled'' states, the latter being defined as follows
\begin{defi} 
A pure state $\Phi$ of system $\rA\otimes \rB$ is \emph{maximally entangled} if no other pure state   of  $\rA\otimes \rB$ is more entangled than $\Phi$, except for the states that are equivalent  to $\Phi$ under  local reversible transformations---i.e.\ if for every   $\Psi \in\Pur\St \left(\rA\otimes \rB\right)$ one has 
$$   \Psi  \succeq_{\rm ent}  \Phi  \quad   \Longrightarrow \quad   
 \Psi    =  \left(  \map   U \otimes \map V\right) \Phi   
   $$
for some reversible transformations $\map U:\rA\to \rA$ and $\map V :  \rB\to \rB$. 
\end{defi} 
 
Theorem~\ref{theo:duality} directly implies   the following.
  \begin{cor}
 The purification of a maximally mixed state is maximally entangled. 
\end{cor}
\Proof     
 Suppose that  $\Psi \in\Pur\St \left(\rA\otimes \rB\right)$ is more entangled than $\Phi$, where $\Phi$ is a purification of the maximally mixed state of system $\rA$ (assuming such a state exists for system $\rA$).   By  theorem~\ref{theo:duality}, the marginal of $\Psi$ on system $\rA$, denoted by $\rho$, must satisfy $\rho\succeq_{\rm mix}  \chi$.   Since $\chi$ is maximally mixed, this implies $\rho  =  \chi$. The uniqueness of purification then implies the condition   $\Psi   =   \left(\map I_\rA\otimes \map V_{\rB}\right)   \Phi$ for some reversible transformation  $\map V_{\rB}$ on $\rB$. \qed  

As noted earlier in the paper, under the standard assumptions of convexity and compactness of the state space, the maximally mixed state is not only a maximal element of the mixedness relation, but also the maximum  [cf.\ Eq.~\eqref{maxmix}].  Similarly, under the same standard assumptions, it is immediate to obtain that the purification of a maximally mixed state  is  more entangled than every   state, namely 
\[
\Phi   \succeq_{\rm ent}   \Sigma  \qquad \forall \, \Sigma \in  \St \left(\rA\otimes \rB\right)  \, .  
\]
The relation follows directly from theorem~\ref{theo:duality} when $\Sigma$ is a pure state and in the general case can be proved by convexity, using the fact that the set of LOCC channels is closed under convex combinations.

\subsection{Duality between information erasure and entanglement generation}

The entanglement-thermodynamics duality establishes a link between the two tasks of erasing information and generating entanglement.    
By \emph{erasing information} we mean resetting a mixed state  to a fixed pure state of the same system \cite{Landauer}.  
Clearly, erasure is  a costly operation in the resource theory of purity: there is no way to transform a non-pure  state into a pure state by using only RaRe channels (cf.\ proposition~\ref{prop:more than pure}).  The dual  operation  in the resource theory of entanglement is the generation of entangled states from product states.    By the duality, the impossibility of erasing information by RaRe channels and the impossibility of generating entanglement by LOCC are one and the same thing. 

The relation between information erasure and entanglement generation suggests that  the cost of erasing a mixed state  $\rho$ could be identified with the cost of generating  the corresponding entangled state  $\Psi$.   For example, one may choose  a fixed entangled state $\Phi$ as a reference ``unit of entanglement'' and ask how many copies of $\Phi$ are needed to generate $\Psi$ through LOCC operations.  The number of entanglement units needed to generate $\Phi$ could then be taken as a measure of the cost of erasing $\rho$.    
We now explore this idea at the heuristic level, discussing first a model of erasure and then connecting it with the generation of entanglement.    
  Suppose that erasure is implemented by \emph{i)} performing a pure measurement, \emph{ii)} writing down the outcome on a classical register, \emph{iii)} conditionally on the outcome, performing a reversible transformation that brings the system to a fixed pure state, and finally \emph{iv)} erasing the classical register. Of course, this model assumes that some systems described by the theory  can act as ``classical registers'', meaning that they have perfectly distinguishable pure states.  Assuming the validity of Landauer's principle at  the classical  level, the cost of erasing the classical register is  then equal to  the Shannon entropy of the outcomes multiplied   by  $k_{\rB} T$, $k_{\rB}$ and $T$ being  the Boltzmann constant and the temperature, respectively \cite{Landauer}.  Minimizing the entropy  over all possible measurements at step \emph{i)}, one would then obtain the  measurement entropy, as defined in Eq.~\eqref{measent}.
  Hence,  the  minimum cost for erasing $\rho$ is given by $k_{\rB} T    H\left(\rho\right)$.      
Note that this heuristic conclusion implicitly assumes that the operations \emph{i-iii)} can be performed for free.   This is the case in quantum theory, where \emph{i)} the measurement attaining minimum Shannon entropy is projective and the overall transformation associated to it is a random unitary channel,    {\em ii)} the measurement outcome can be written down via a unitary operation on the system and the classical register, and {\em iii)}   the state of the system can be reset via another joint unitary operation.    In physical theories other than quantum  and classical theories, however, the request that the operations \emph{i-iii)} are free  is non-trivial and would need to be further analysed in terms of physical axioms.   

Suppose now that we want to erase an \emph{unknown} state $\rho$.  Since the state is unknown, the relevant quantity here is  the \emph{worst case cost of erasure}, defined as the supremum of the cost over all possible states.   Since the measurement entropy is monotone under the mixedness relation, in finite dimensions the supremum is attained for the maximally mixed state $\chi$, so that the worst case cost of erasure is given by $k_{\rB} T  H\left(\chi\right)$.    
  This result allows us to make an interesting connection with the work  by Brunner \emph{et al}    \cite{Brunner-thermo}, who considered the task of erasure in general probabilistic theories.  
   Specifically, they considered  the number of states that can be perfectly distinguished by a measurement and adopted the  logarithm of this  number   as a  measure of the cost of erasure.  In their analysis they   considered probabilistic quantum and classical  theories, as well as alternative theories  with hypercube state spaces, wherein measurements can distinguish at most two states.   In all these theories the logarithm of the dimension is exactly equal to the measurement entropy of the maximally mixed state.      Thanks to this fact, erasure cost defined  in Ref.     \cite{Brunner-thermo}   coincides with the worst case erasure cost defined above.  It is an open question  whether the two definitions coincide in all canonical theories of purity, and, if not, which conditions are needed for the two definitions to coincide.  
 
Let us now look at erasure from the dual point of view.   Since the duality inverts the order, the dual of an erasure protocol consisting of operations {\em i),i),iii)} and {\em  iv)} will be an entanglement generation protocol consisting of the dual operations in the opposite order {\em iv),iii),ii)} and {\em i)}.   The dual of \emph{iv)} is an operation that generates a purification of the classical register.  The duals of the free operations \emph{i-iii)} are LOCC operations that convert the initial entangled state into the state $\Psi$.     Now, by the duality we can measure the cost of generating $\Psi$ in terms of the measurement entropy $H\left(\rho\right)$.    But  what is the operational meaning of this choice?  Again, the duality suggests an  answer.  Classically, the Shannon entropy can be interpreted as the asymptotic rate at which random bits can be extracted from a given probability distribution.  Dually, the inverse relation must hold between the purifications: Referring to the purification of a random bit as an \emph{ebit}, we have that the Shannon entropy is  the number of ebits needed asymptotically to generate the purification of a given probability distribution by LOCC.          Minimizing over all probability distributions one can characterize  the measurement entropy as the minimum number of ebits needed  to  asymptotically generate the state $\Psi$ by LOCC.      Although partly based on heuristics, the argument provides already a good illustration of the far reaching consequences of the entanglement-thermodynamics duality, which allowed to identify the cost of erasing a state with the number of ebits needed to generate its purification.

\subsection{Entropy sinks and entanglement reservoirs}

Let us consider now the task of   erasure assisted by a catalyst, namely  a system $\rC$ whose state  remains unaffected by the erasure operation.  In this case, the operation of erasure  transforms the product state $\rho\otimes \gamma   \in\St \left(\rA\otimes \rC\right)$  into the state $\alpha_0\otimes \gamma$ for some pure state $\alpha_0 \in\Pur\St \left(\rA\right)$.   
 By duality, it is immediate to see that catalyst-assisted erasure is equivalent to catalyst-assisted entanglement generation: 
\begin{cor}  
Let $\Psi$ and $\Gamma$ be two pure states of systems $\rA\otimes \rB$ and $\rC\otimes \rD$, respectively, and let $\rho$ and $\gamma$ be their marginals on systems $\rA$ and $\rC$, respectively.  Then, the following are equivalent
\begin{enumerate}
\item $\rho$ can be erased by a RaRe channel using $\gamma$ as a catalyst
\item  $\Psi$ can be generated by a LOCC channel using $\Gamma$ as a catalyst.      
\end{enumerate}
\end{cor}
If such catalysts existed, they  would behave like ``entropy sinks'', which  absorb mixed states without becoming more mixed, or like ``entanglement reservoirs'',  from which entanglement can be borrowed indefinitely.  For example,  suppose that $\Psi$ can be generated freely using $\Gamma$ as a catalyst.  Then, every measure of pure state entanglement $E$ that    it is additive on product states would have to satisfy the relation 
\[
E  \left(\Gamma\right)   \ge    E\left(\Psi\right)  +  E\left(\Gamma\right)\,  . \] 
Assuming that the measure assigns a strictly positive value to every entangled state. the above relation can only be satisfied if $E\left(\Gamma\right)  =  + \infty$. In other words,  the catalyst's state must be  infinitely entangled. 
It is then natural to ask whether the impossibility of infinitely entangled/infinitely mixed states  follow from our axioms.    The answer is affirmative in the finite-dimensional case, but counterexamples exist in infinite dimensions.  
For the finite-dimensional case, we have the following 
\begin{prop}\label{prop:finitecat}
Let $\rA \otimes \rC$ be  a finite system. Then, it is impossible to erase a mixed state of $\rA$  using $\rC$ as a catalyst.  
\end{prop}
The proof is presented in appendix~\ref{app:finitecat}.    

In the infinite-dimensional case, a heuristic  counterexample is as follows:  imagine a scenario where  system $\rC$ consists of an infinite chain of identical systems, with each system in the chain equivalent to  $\rA$, namely $\rC  =  \bigotimes_{i\in\mathbb Z}  \,   \rA_i$,  $\rA_i  \simeq \rA$.  Loosely speaking, we may choose  the state $\gamma$ to be the product state $\gamma  =  \gamma_L\otimes \gamma_R$, where $\gamma_L$ is a state on the left side of the chain, consisting of infinite copies of the pure state $\alpha_0$, and $\gamma_R$ is  a state on the right of the chain, consisting of infinite copies of the  mixed state $\rho$.  It is then natural to expect that the state $\rho\otimes \gamma$ can be reversibly transformed into the state $\alpha_0\otimes \gamma$, simply by swapping system $\rA$ with the first system on the left of the chain and subsequently shifting the whole chain by one place to the right.   

The above counterexample is heuristic, because the notion of infinite tensor product is not defined in our formalism.  However, infinite tensor products can be treated rigorously, at least in the quantum case, and the intuition of our counterexample turns out to be correct.  In the dual task of catalytic entanglement generation, the rigorous version of this argument was presented by Keyl, Matsui, Schlingemann, and Werner \cite{Werner}, who exhibited an example of infinite spin chain from which arbitrarily large amounts of entanglement can be generated for free.

\section{\label{sec:symmetric purification} Symmetric Purification}

As we observed in the previous section, the ability to erase information/generate entanglement for free has  undesirable consequences for the resource theories of purity and entanglement.   
 These scenarios can be excluded at the level of first principles, by postulating the following
\begin{ax}[
No Entropy Sinks]\label{ax:nores}
Random reversible dynamics cannot achieve erasure, even with the assistance of a catalyst. 
\end{ax}
In addition to being a requirement for a sensible resource theory of entanglement,   
Axiom~\ref{ax:nores} has a surprising twist: in the context of the other axioms, it implies  that Local Exchangeability is equivalent to the existence of  \emph{symmetric} purifications, defined as follows 
\begin{defi}
Let $\rho$ be a state of system $\rA$ and let $\Psi$ be a pure state of $\rA\otimes \rA$.  We say that $\Psi$ is a \emph{symmetric purification} of $\rho$ if 
\begin{align*}
\begin{aligned}\Qcircuit @C=1em @R=.7em @!R { & \multiprepareC{1}{\Psi}    & \qw \poloFantasmaCn{\rA} &  \qw  &  \qquad &=   \qquad    &      \prepareC {\rho}  & \qw \poloFantasmaCn{\rA}  &  \qw   \\  & \pureghost{\Psi}    & \qw \poloFantasmaCn{\rA}  &   \measureD{\Tr}  &&&& }\end{aligned}   
\end{align*}  
and  
\begin{align*}
\begin{aligned}\Qcircuit @C=1em @R=.7em @!R { 
& \pureghost{\Psi}    & \qw \poloFantasmaCn{\rA}  &   \measureD{\Tr}   &&&&  \\
& \multiprepareC{-1}{\Psi}    & \qw \poloFantasmaCn{\rA} &  \qw    &    \qquad &    =     \qquad &      \prepareC {\rho}  & \qw \poloFantasmaCn{\rA}  &  \qw   
 \quad . }    \end{aligned}  
\end{align*}
\end{defi} 
This definition leads  us to  an upgraded version of the Purification axiom:

\begin{ax}[Symmetric Purification]
Every state has a symmetric purification. Every purification is essentially unique. 
\end{ax}
The key result is then given by the following:  

\begin{theo}\label{theo:symmetric}
In a causal theory satisfying Purity Preservation and No Entropy Sinks, the following axioms are equivalent:  
\begin{enumerate}
\item  Local Exchangeability and  Purification 
\item  Symmetric Purification.
\end{enumerate} 
\end{theo}   
The proof is presented in appendix~\ref{app:symmetric}.   

This result identifies Purity Preservation and Symmetric Purification as the key axioms at the foundation of the entanglement-thermodynamics duality and, ultimately, as  strong candidates for a reconstruction of quantum thermodynamics from first principles.  
We stress that the axiom No Entropy Sinks  is needed only for infinite-dimensional systems, while for finite dimensional systems its validity can be proved (cf.\ proposition~\ref{prop:finitecat}).  

One of the bonuses of Symmetric Purification is that the marginals of a pure state are ``equivalent'', in the following sense:  
\begin{prop}\label{prop:marginals}
Let $\Psi$ be a pure state of system $\rA\otimes \rB$ and let $\rho_\rA$ and $\rho_\rB$ be its marginals on systems $\rA$ and $\rB$, respectively.  Then, one has  
\[
\begin{aligned}\Qcircuit @C=1em @R=.7em @!R {    & \prepareC{\rho_\rB}   &  \qw \poloFantasmaCn{\rB}  &  \qw   \\
    & \prepareC{\alpha}   &  \qw \poloFantasmaCn{\rA}  &  \qw }     
\end{aligned}~
= \!\!\!\!
\begin{aligned}\Qcircuit @C=1em @R=.7em @!R {    & \prepareC{\rho_\rA}   &  \qw \poloFantasmaCn{\rA}  &\multigate{1}{\map U}  &  \qw \poloFantasmaCn{\rB}  &  \qw   \\
    & \prepareC{\beta}   &  \qw \poloFantasmaCn{\rB}  &  \ghost{\map U}  &  \qw  \poloFantasmaCn{\rA}   & \qw }     
\end{aligned} ~ ,
\]
where $\alpha$ and $\beta$ are pure states of $\rA$ and $\rB$, respectively, and $\map U$ is a reversible transformation.  
\end{prop}
The proof can be found in appendix~\ref{app:marginals}.  As a consequence of this result, we have that the states $\rho_\rA\otimes\beta$ and $\alpha\otimes 
\rho_\rB$ have the same purity, for every possible purity monotone.   Equivalently, we can say that the two marginal states have the same mixedness, for every measure of mixedness.  
 
\section{\label{sec:Conclusions}Conclusions}

While entanglement is not a uniquely quantum feature, the remarkable ways in which it is intertwined with thermodynamics appear to be far more specific.   
Understanding these links  at the level of basic principles is expected to reveal new clues to  the foundations of  quantum theory, as well as to the foundations of  thermodynamics.    
With this motivation in mind, we set out to search for  the roots of the relation between entanglement and entropy, adopting  an operational, theory-independent approach. 
We attacked the problem from what is arguably the most primitive link:    the duality between the resource theory of entanglement (where the free transformations are those achievable by two spatially separated agents via local operations and classical communication)  and the resource theory of purity (where the free  transformations are those achievable by an agent who has limited control on the dynamics of the system).  
By the duality, every free operation  in the resource theory of purity admits an equivalent description as  a free operation in the resource theory of  entanglement.     The duality leads to an identification between measures of mixedness (i.e.\ lack of purity) and measures of pure bipartite entanglement.   Under suitable conditions, the latter can be extended to measures of mixed-state entanglement.

Let us elaborate on the implications of our results.   Our reconstruction of the entanglement-thermodynamics duality hints at a simple, physically motivated idea: the idea that nature should  admit a fundamental level of description where all states  are pure, all dynamics reversible, and all measurements pure.  
Two of our axioms clearly express this requirement:  \emph{i)}    Purification is equivalent to the existence of a pure and reversible level of description for states and channels, and  \emph{ii)}  Purity Preservation ensures that  such a description remains consistent when different, possibly non-deterministic processes are connected. 
The remaining axiom,  Local Exchangeability, appeared to be slightly more mysterious at first sight. Nevertheless, the duality clarified its significance:     
 for every finite system (and more generally, for every system where all states have finite entanglement), Local Exchangeability is equivalent to the existence of \emph{symmetric} purifications---that is, purifications where the purifying system is a twin of the purified system.   In summary, all the axioms used to derive the duality are requirements about the possibility to come up with an ideal description of the world, satisfying simple requirements of purity, reversibility, and symmetry. 
 
A natural question is whether these axioms single out quantum theory.  Strictly speaking, the answer cannot be affirmative, because all our axioms are satisfied by also by the variant of quantum theory based on real Hilbert spaces \cite{stueckelberg,wootters1990local}.  Hence, the actual question is whether real and complex quantum theory are the only two examples of theories satisfying the axioms. 
While an affirmative answer  is logically possible, we do not expect it to be the case.  The reason is that our axioms do not place any restriction on measurements:  for example, our proof of the duality does not require one to assume an operational analogue of Naimark's theorem, stating that every measurement can be implemented as an ideal measurement at the fundamental level. 
Overall, in the general purification philosophy of our work,   it is natural to expect that full characterization of quantum theory will require at least one requirement about the existence of a class of ideal  measurements that generalize projective quantum measurements.   

Naimark-type axioms for measurements has been recently put forward by one of the authors \cite{Chiribella-Yuan2014,Chiribella-Yuan2015}, for the purpose of  deriving bounds on quantum nonlocality and contextuality.  A natural development of our work  is to investigate the consequences of these axioms on the entanglement-thermodynamics duality.   From such development, we expect a solution to most of the outstanding questions arising from the present paper.  Among them, an important   one concerns the asymptotic limit of many identical copies:  in quantum theory, it is well known that asymptotically there exists a unique measure of pure bipartite entanglement \cite{Entanglement-concentration,Popescu-entanglement,Vedral-entanglement}---namely, the von Neumann entropy.  Under which conditions does this result hold in the general probabilistic scenario? 
In order to address the question, the most promising route is to add an axiom about ideal measurements, which, combined with Purification and Purity Preservation, guarantees that mixed states can be ``diagonalized'', that is, decomposed as random mixtures of perfectly distinguishable pure states \cite{QPL15}.  The consequences of this diagonalization result for the entanglement-thermodynamics duality will be discussed in a forthcoming paper \cite{Chiribella-Scandolo-15-2}.

Another open question concerns the physical interpretation of the duality.   So far, the duality has been presented as a one-to-one correspondence between two operational scenarios, one  involving a single agent with limited control and the other involving two spatially separated agents performing LOCC operations on a pure state. Inspired by the paradigm of the ``fundamentally pure and reversible description'', one may be tempted to regard the pure-state side of the duality as more fundamental.  To push   this idea further, one would have to consider a completely-coherent version of the LOCC operations, where Alice's and Bob's operations are replaced by control-reversible channels \cite{Harrow}.  Restricting the global dynamics of composite systems to these completely-coherent evolutions  appears as a promising direction in the programme of deriving effective thermodynamic features  from the reversible dynamics of a composite system  \cite{Popescu-Short-Winter,Lubkin,Gemmer-Otte-Mahler,Canonical-typicality,Mahler-book,Concentration-measure,BrandaoQIP2015,Muller-blackhole}.  
While it is early to predict all the applications of the completely-coherent paradigm, our work provides the basic theoretical framework and motivation to embark in this new exploration.

\acknowledgments 

We are grateful to M Piani for a useful discussion during QIP 2015 and to M M\"uller for drawing our attention to the notion of group majorization in Ref.~\cite{Muller3D}.    We are also grateful to the anonymous referee of NJP who suggested to add a discussion on the cost of erasure.   This work is supported  by the Foundational Questions Institute through the large grant ``The fundamental principles of information dynamics'' (FQXi-RFP3-1325), by the National Natural Science Foundation of China through Grants 11450110096, 11350110207, and by the 1000 Youth Fellowship Program of China. 
GC acknowledges the hospitality of the Simons Center for the Theory of Computation and of Perimeter Institute, where part of this work was done.  Research at Perimeter Institute for Theoretical Physics is supported in part by the Government of Canada through NSERC and by the Province of Ontario through MRI. The research by CMS has been supported by a scholarship from ``Fondazione Ing.\ Aldo Gini'' and by the Chinese Government Scholarship.

\appendix

\section{Proof of proposition~\ref{prop:pure states reversible}}\label{app:proof_reversible}
Suppose that $\mathcal{U}\psi$ can be written as a coarse-graining as follows
\begin{equation}
\mathcal{U}\psi=\sum_{i}\rho_{i}   \, . \label{eq:pure reversible}
\end{equation}
To prove that the state $\map U  \psi$ is pure, now we show that the refinement $\left\{\rho_i\right\}$ is trivial.  
Indeed, by applying $\mathcal{U}^{-1}$ to both sides of Eq.~\eqref{eq:pure reversible},  we obtain 
\[
\psi=\sum_{i}\mathcal{U}^{-1}\rho_{i} \, .
\]
Since $\psi$ is pure, this implies that $\map U^{-1}  \rho_i  =   p_i  \,  \psi$ for some probability distribution $\left\{p_i\right\}$.   Hence, by applying $\map U$ on both sides, we  obtain $\rho_i  =  p_i \map U \psi$.     This concludes the proof that $\{\rho_i\}$ is a trivial refinement of $\map U\psi$ and, therefore, that $\map U\psi$ is pure.   The converse can be proved in the same way by applying the reversible channel
$\mathcal{U}^{-1}$ to $\mathcal{U}\psi$.  \qed
\section{Proof of proposition \ref{prop:canonical}} \label{app:canonical}

 $1\Longrightarrow 2$. If the theory is canonical,  every pure state  $\psi  \in  \Pur\St \left(\rA\right)$ is comparable to every pure state  $\varphi  \in \Pur\St \left(\rA\right)$. Suppose, for instance that   $\psi$ is more controllable than $\varphi$. Then, by proposition~\ref{prop:more than pure}, there exists a reversible channel $\mathcal{U}$ such that $\psi = \mathcal{U}\varphi$, thus showing that the group of reversible transformations acts transitively on the set of pure states.

  $2\Longrightarrow 3$.    Every state $\rho$ can be expressed as a convex combination of the form
$\rho=\sum_{i}p_{i}\varphi_{i}$, where $\left\{p_i\right\}$ is a probability distribution allowed by the theory and $\varphi_i$ are pure states. Now, suppose that  $\psi$ is a pure state. For every $i$, by picking a reversible channel $\mathcal{U}^{\left(i\right)}$
such that $\mathcal{U}^{\left(i\right)}\psi=\varphi_{i}$, one obtains the relation $\rho=\sum_{i}p_{i}\mathcal{U}^{\left(i\right)}\psi$, meaning that $\psi$  is more  controllable than  $\rho$.  Since $\rho$ is generic, we conclude that $\psi$ is more controllable than every state.  
  
$3\Longrightarrow 1$.  
Suppose there exists a state $\rho$ that is more controllable than every  state.  Specifically, $\rho$ must be more controllable than every pure state $\psi$.   By proposition~\ref{prop:more than pure}, $\rho$ must be pure and there exists a reversible transformation $\map U$ such that $\rho  =  \map U  \psi$. This shows that $\psi$ is more controllable than $\rho$, which, in turn is more controllable than any state. Hence $\psi$ is more controllable than every state, and, specifically, more controllable than every pure state. Since $\psi$ is generic, the theory is canonical. \qed

\section{Proof of corollary~\ref{cor:LU}}\label{app:LU}

Clearly, if $\Psi$ and $\Psi'$   are equivalent under local reversible transformations, then they are equally entangled.  To prove the converse, note that,  by the duality, the marginals of $\Psi$ and $\Psi'$ on system $\rA$, denoted by $\rho$ and $\rho'$, are equally mixed.    Since $\rA$ is finite-dimensional, this implies $\rho' =   \map U  \rho$ for some reversible transformation $\map U$.  As a consequence, $   \Psi'$  and $\left(\map U\otimes \map I_\rB\right)  \Psi$ are two purifications of $\rho'$.  By the essential uniqueness of purification, we then have $\Psi'   =   \left( \map U  \otimes \map V \right)  \Psi$ for some reversible transformation $\map V$.  \qed

\section{Purity monotones and Schur-convex functions} \label{app:schur} 

In classical probability theory,  states  are probability distributions  over finite sets   and reversible transformations are permutation matrices, of the form $  \Pi_{mn}   =   \delta_{m,  \pi  (n)}$ where $\pi$ is a permutation.   According to  definition~\ref{def:mixedness relation},  a state $ \bs  p  =  \left(\begin{array}{ccc}
p_{1} & \ldots & p_{n}\end{array}\right)^{T}$  is purer than another state $\bs p'  =  \left(\begin{array}{ccc}
p'_{1} & \ldots & p'_{n}\end{array}\right)^{T}$ if    
\[   \bs p'    =     \sum_i     q_i    \Pi_i   \bs p  \, ,  \]  
where $\left\{q_i\right\}$ are probabilities and $\left\{\Pi_i\right\}$ are permutation matrices.   This notion is equivalent to the classical notion of majorization:  in short, $\bs p$ is purer than $\bs p'$  if and only if the vector $\bs p'$ is majorized by the vector $\bs p$.  
Hence,  a function $P:  \R^n \to \R$  is a purity monotone iff it is a Schur-convex function.

The parallel between purity monotones and Schur-convex functions continues with proposition \ref{prop:monotones}.   In classical probability theory, a  function       $P:  \R^n \to \R$ is  \emph{symmetric}  if $  P\left(\bs x\right)  =   P\left(  \Pi   \bs x \right)$ for every permutation matrix $\Pi$.  
A well-known result is that every convex symmetric function is Schur-convex \cite{olkin}.  Our proposition \ref{prop:monotones} is the operational version of this statement:  every convex function $P  :  \St  \left(\rA\right)  \to \R$ satisfying the condition $P\left(\rho\right)  =  P\left(  \map U  \rho\right)$ for every reversible transformation $\map U$ is a purity monotone.  
The proof is elementary.   Suppose that $\rho$ is purer than $\rho'$, namely  $\rho'  =  \sum_i   p_i    \map U_i   \rho$.   Then, one has 
\begin{align*}
P\left(  \rho'\right)  &  \le   \sum_i  p_i    P\left(\map U_i  \rho \right)  \\
   &   =  \sum_i  p_i    P\left(\rho\right)  \\
   &  =  P\left(\rho\right)  \, ,
\end{align*}
having used convexity in the first inequality and invariance in the first equality.  In the classical case, this (trivial) proof provides a simpler proof of the well-known result for convex symmetric functions (cf.\ C.2 of Ref.~\cite{olkin}).

\section{Proof of proposition~\ref{prop:finitecat}}\label{app:finitecat}  

Let us prove the contrapositive:  if a state can be erased using system $\rC$ as a catalyst, then the state must be pure.    Specifically, suppose that $\rho \in\St \left(\rA\right)$ can be erased, with the catalyst in the state $\gamma  \in  \St\left(\rC\right)$.  By definition, this means that $\rho  \otimes  \gamma  \preceq_{\rm mix}   \alpha_0  \otimes \gamma$ for some pure state $\alpha_0$.   On the other hand, one has $\rho  \succeq_{\rm mix}   \alpha_0  $, which implies    $\rho  \otimes  \gamma  \succeq_{\rm mix}   \alpha_0  \otimes \gamma$---hence  $\rho  \otimes  \gamma$ and $\alpha_0  \otimes  \gamma$ are equally mixed.     Since $\rA\otimes \rC$ is a finite system, this means that there exists a reversible transformation $\map U$ such that 
\begin{equation}\label{aaaaa}
\map U  \left(\alpha_0  \otimes  \gamma \right)   =  \rho  \otimes  \gamma \, ,
\end{equation}
[cf.\ Eq.~\eqref{reveq}].     Now, let us choose a basis for $\St_\R \left(\rA\otimes \rC \right)$, such that the reversible transformations are represented by orthogonal matrices.  Following Ref.~\cite{Muller3D}, we  consider the Schatten 2-norm associated with this basis, defined as  \[\left\Vert   v   \right\Vert_2   : =    \sqrt{  \sum_{i=1}^{D_{\rA\otimes \rC}}   v_i^2} \, , \] 
where $v $ is a generic element of the vector space $\St_\R \left(\rA\otimes \rC\right)$ and $\left(v_i\right)_{i=1}^{D_{\rA\otimes \rC}}$  are the expansion coefficients of $v$.  With this definition, we have the relation
\begin{align*}
\left\Vert     \alpha_0  \otimes  \gamma   \right\Vert_2    &  =  \left\Vert    \map U \left( \alpha_0  \otimes  \gamma  \right) \right\Vert_2  \\
 &  = \left\Vert     \rho  \otimes  \gamma   \right\Vert_2   \\
&   \le \sum_i      p_i   \left\Vert    \alpha_i  \otimes \gamma  \right\Vert_2   \\
&  =    \left\Vert  \alpha_0  \otimes \gamma  \right\Vert_2  \, ,
\end{align*}
the first and fourth lines following from the invariance of the $2$-norm under orthogonal transformations, the second line following from Eq.~\eqref{aaaaa}, and the third line following from the triangular inequality, having chosen a convex decomposition of $\rho$ as $\rho  =  \sum_i  p_i \alpha_i$ for suitable pure states $\left\{\alpha_i\right\}$.   In conclusion, we must have the equality 
\[  \left\Vert  \sum_i    p_i   \left (  \alpha_i  \otimes  \gamma \right)    \right\Vert_2   =   \sum_i      p_i   \left\Vert    \alpha_i  \otimes \gamma  \right\Vert_2    \, .\]
In order for this to be possible, all the terms $\alpha_i \otimes \gamma$ must be proportional to one another: in other words, $\rho$ must be pure.  \qed 

\section{Proof of theorem~\ref{theo:symmetric}}\label{app:symmetric}  

Let us show that the first set of axioms (Purity Preservation, Local Exchangeability, Purification, and No Entropy  Sinks) implies the second (Purity Preservation, Symmetric Purification, No Entropy Sinks).   To this purpose, it is sufficient to show that every state has a symmetric purification.  This can be done as follows: 

 Let $\rho$ be a state of system $\rA$, and let $\Psi \in \Pur\St \left(\rA \otimes \rB \right)$  be one of its purifications. By Local Exchangeability there exist two channels $\map C$ and $\map D$ such that    
\begin{align*}
\begin{aligned}\Qcircuit @C=1em @R=.7em @!R { & \multiprepareC{1}{\Psi}    & \qw \poloFantasmaCn{\rA} &\gate{\cC} & \qw \poloFantasmaCn{\rB} &\qw \\  & \pureghost{\Psi}    & \qw \poloFantasmaCn{\rB}  &\gate{\cD} & \qw \poloFantasmaCn{\rA} &\qw}\end{aligned} ~=\!\!\!\! \begin{aligned}\Qcircuit @C=1em @R=.7em @!R { & \multiprepareC{1}{\Psi}    & \qw \poloFantasmaCn{\rA} &   \multigate{1}{  {\tt  SWAP}}   &  \qw  \poloFantasmaCn{\rB}  &  \qw   \\  & \pureghost{\Psi}    & \qw \poloFantasmaCn{\rB}   &   \ghost{  {\tt  SWAP}}   &    \qw \poloFantasmaCn{\rA} &\qw  }   \end{aligned}    
\end{align*}
Now, in a theory satisfying Purification, every channel can be realized through a reversible transformation acting on the system and on an environment, initially in a pure state and finally discarded \cite{Chiribella-purification}. Specifically, channel $\map C$ can be realized as  
\begin{align*}
\begin{aligned}
\Qcircuit @C=1em @R=.7em @!R {&\qw \poloFantasmaCn{\rA}  & \gate{{\cC}}  &\qw  \poloFantasmaCn{\rB}  &\qw} 
\end{aligned}  ~ : = \!\!\!\!
\begin{aligned}
\Qcircuit @C=1em @R=.7em @!R { 
&\prepareC{\eta} & \qw \poloFantasmaCn{\rE} &\ghost{\cU} &\qw \poloFantasmaCn{\rE'} &\measureD{\Tr}\\ 
& & \qw \poloFantasmaCn{\rA}  &\multigate{-1}{\cU} &\qw \poloFantasmaCn{\rB} &\qw &  }   
\end{aligned} ~ ,
\end{align*}
where  $\rE$ and $\rE'$ are suitable systems, $\map U$ is a reversible transformation, and $\eta $ is a pure state. 
Similarly, channel $\map D$ can be realized as 
\begin{equation}\label{D}
\begin{aligned}
\Qcircuit @C=1em @R=.7em @!R {&\qw \poloFantasmaCn{\rB}  & \gate{\cD}  &\qw  \poloFantasmaCn{\rA}  &\qw} 
\end{aligned}  ~ : = \!\!\!\!
\begin{aligned}
\Qcircuit @C=1em @R=.7em @!R { 
& & \qw \poloFantasmaCn{\rB}  &\multigate{1}{\map V} &\qw \poloFantasmaCn{\rA} &\qw &   \\
&\prepareC{\varphi} & \qw \poloFantasmaCn{\rF} &\ghost{\map V} &\qw \poloFantasmaCn{\rF'} &\measureD{\Tr}\\ 
 }   
\end{aligned}  ~ .
\end{equation}
Inserting the realizations of $\map C$ and $\map D$ in the local exchangeability condition, we obtain  
\begin{align*}
\begin{aligned}\Qcircuit @C=1em @R=.7em @!R { 
&\prepareC{\eta} & \qw \poloFantasmaCn{\rE} &\ghost{\cU} &\qw \poloFantasmaCn{\rE'} &\measureD{\Tr} \\
& \multiprepareC{1}{\Psi}    & \qw \poloFantasmaCn{\rA} &\multigate{-1}{\cU} & \qw \poloFantasmaCn{\rB} &\qw \\  & \pureghost{\Psi}    & \qw \poloFantasmaCn{\rB}  &\multigate{1}{\map V} & \qw \poloFantasmaCn{\rA} &\qw  \\
&\prepareC{\varphi} & \qw \poloFantasmaCn{\rF} &\ghost{\map V} &\qw \poloFantasmaCn{\rF'} &\measureD{\Tr}  }\end{aligned} ~=\!\!\!\! \begin{aligned}\Qcircuit @C=1em @R=.7em @!R { & \multiprepareC{1}{\Psi}    & \qw \poloFantasmaCn{\rA} &   \multigate{1}{  {\tt  SWAP}}   &  \qw  \poloFantasmaCn{\rB}  &  \qw   \\  & \pureghost{\Psi}    & \qw \poloFantasmaCn{\rB}   &   \ghost{  {\tt  SWAP}}   &    \qw \poloFantasmaCn{\rA} &\qw  }   \end{aligned}    
~ .
\end{align*}
Since the pure state on the l.h.s.\ is the purification of a pure state, by proposition~\ref{prop:pure-product},  it must be of the product form
\begin{align*}
\begin{aligned}\Qcircuit @C=1em @R=.7em @!R { 
&\prepareC{\eta} & \qw \poloFantasmaCn{\rE} &\ghost{\cU} &\qw \poloFantasmaCn{\rE'} &\qw \\
& \multiprepareC{1}{\Psi}    & \qw \poloFantasmaCn{\rA} &\multigate{-1}{\cU} & \qw \poloFantasmaCn{\rB} &\qw \\  & \pureghost{\Psi}    & \qw \poloFantasmaCn{\rB}  &\multigate{1}{\map V} & \qw \poloFantasmaCn{\rA} &\qw  \\
&\prepareC{\varphi} & \qw \poloFantasmaCn{\rF} &\ghost{\map V} &\qw \poloFantasmaCn{\rF'} &\qw  }\end{aligned} 
~=\!\!\!\! \begin{aligned}\Qcircuit @C=1em @R=.7em @!R { 
&  \multiprepareC{3}{\Gamma}  &  \qw \poloFantasmaCn{\rE'}   &  \qw   &\qw &\qw  & \qw &\qw  \\
&\pureghost{\Gamma}& & \multiprepareC{1}{\Psi}    & \qw \poloFantasmaCn{\rA} &   \multigate{1}{  {\tt  SWAP}}   &  \qw  \poloFantasmaCn{\rB}  &  \qw   \\  
&\pureghost{\Gamma}  & & \pureghost{\Psi}    & \qw \poloFantasmaCn{\rB}   &   \ghost{  {\tt  SWAP}}   &    \qw \poloFantasmaCn{\rA} &\qw \\
 &  \pureghost{\Gamma}  &  \qw \poloFantasmaCn{\rF'}   &  \qw   &\qw &\qw  & \qw &\qw   }   \end{aligned}    
\end{align*}  
for some pure state $\Gamma$.  
The above equation shows that  the state $\Gamma $ can be generated by LOCC using $\Psi$  as a catalyst.  By the No Entropy Sinks requirement, we have that $\Gamma$ must be a product state, i.e.\ $\Gamma  =  \eta'\otimes \varphi'$ for two pure states $\eta'$ and $\varphi'$.    Hence, the local exchangeability condition becomes
\begin{align*}
\begin{aligned}\Qcircuit @C=1em @R=.7em @!R { 
&\prepareC{\eta} & \qw \poloFantasmaCn{\rE} &\ghost{\cU} &\qw \poloFantasmaCn{\rE'} &\qw \\
& \multiprepareC{1}{\Psi}    & \qw \poloFantasmaCn{\rA} &\multigate{-1}{\cU} & \qw \poloFantasmaCn{\rB} &\qw \\  & \pureghost{\Psi}    & \qw \poloFantasmaCn{\rB}  &\multigate{1}{\map V} & \qw \poloFantasmaCn{\rA} &\qw  \\
&\prepareC{\varphi} & \qw \poloFantasmaCn{\rF} &\ghost{\map V} &\qw \poloFantasmaCn{\rF'} &\qw  }\end{aligned} 
~=\!\!\!\! \begin{aligned}\Qcircuit @C=1em @R=.7em @!R { 
 &  \prepareC{\eta'}  &  \qw \poloFantasmaCn{\rE'}   &  \qw   &\qw &\qw    \\
& \multiprepareC{1}{\Psi}    & \qw \poloFantasmaCn{\rA} &   \multigate{1}{  {\tt  SWAP}}   &  \qw  \poloFantasmaCn{\rB}  &  \qw   \\  
& \pureghost{\Psi}    & \qw \poloFantasmaCn{\rB}   &   \ghost{  {\tt  SWAP}}   &    \qw \poloFantasmaCn{\rA} &\qw \\
 &  \prepareC{\varphi'}  &  \qw \poloFantasmaCn{\rF'}   &  \qw   &\qw &\qw   }   \end{aligned}    
\end{align*}  
or, equivalently, 
\begin{align*}
\begin{aligned}\Qcircuit @C=1em @R=.7em @!R { 
&\prepareC{\eta} & \qw \poloFantasmaCn{\rE} &\qw &\qw &\qw \\
& \multiprepareC{1}{\Psi}    & \qw \poloFantasmaCn{\rA} & \qw &\qw &\qw \\  & \pureghost{\Psi}    & \qw \poloFantasmaCn{\rB}  &\multigate{1}{\map V} & \qw \poloFantasmaCn{\rA} &\qw  \\
&\prepareC{\varphi} & \qw \poloFantasmaCn{\rF} &\ghost{\map V} &\qw \poloFantasmaCn{\rF'} &\qw  }\end{aligned} 
~=\!\!\!\! \begin{aligned}\Qcircuit @C=1em @R=.7em @!R { 
 &  \prepareC{\eta'}  &  \qw \poloFantasmaCn{\rE'}   &\qw  &  \qw   &\ghost{\map U^{-1}} &\qw   \poloFantasmaCn{\rE} &\qw \\
& \multiprepareC{1}{\Psi}    & \qw \poloFantasmaCn{\rA} &   \multigate{1}{  {\tt  SWAP}}   &  \qw  \poloFantasmaCn{\rB} &\multigate{-1}{\map U^{-1}} &  \qw   \poloFantasmaCn{\rA} &\qw \\  
& \pureghost{\Psi}    & \qw \poloFantasmaCn{\rB}   &   \ghost{  {\tt  SWAP}}   &    \qw \poloFantasmaCn{\rA} &\qw  &   \qw     &\qw   \\
 &  \prepareC{\varphi'}  &  \qw \poloFantasmaCn{\rF'}   &  \qw &\qw     &\qw &\qw    &\qw      }   \end{aligned}    
\end{align*}  
Discarding system $\rE$ one obtains  
\begin{align*}
\begin{aligned}\Qcircuit @C=1em @R=.7em @!R {
& \multiprepareC{1}{\Psi}    & \qw \poloFantasmaCn{\rA} & \qw &\qw &\qw \\  & \pureghost{\Psi}    & \qw \poloFantasmaCn{\rB}  &\multigate{1}{\map V} & \qw \poloFantasmaCn{\rA} &\qw  \\
&\prepareC{\varphi} & \qw \poloFantasmaCn{\rF} &\ghost{\map V} &\qw \poloFantasmaCn{\rF'} &\qw  }\end{aligned}
~=\!\!\!\! \begin{aligned}\Qcircuit @C=1em @R=.7em @!R {
 &  \multiprepareC{1}{\Sigma}  &  \qw \poloFantasmaCn{\rA}   &  \qw \\
 &  \pureghost{\Sigma}  &  \qw \poloFantasmaCn{\rA}   &  \qw \\ &  \prepareC{\varphi'}  &  \qw \poloFantasmaCn{\rF'}   &  \qw   }   \end{aligned}      
\end{align*}  
for some suitable state $\Sigma$.   Since the l.h.s.\ is a pure state, $\Sigma$ must be a pure state. Now,  discarding  system $\rF'$ and the second copy of system $\rA$, and recalling Eq.~\eqref{D}, we have
\[
\begin{aligned}\Qcircuit @C=1em @R=.7em @!R {
& \multiprepareC{1}{\Psi}    & \qw \poloFantasmaCn{\rA} & \qw &\qw &\qw \\  & \pureghost{\Psi}    & \qw \poloFantasmaCn{\rB}  &\gate{\map D} & \qw \poloFantasmaCn{\rA} &\measureD{\Tr} }\end{aligned}
~=\!\!\!\! \begin{aligned}\Qcircuit @C=1em @R=.7em @!R {
 &  \multiprepareC{1}{\Sigma}  &  \qw \poloFantasmaCn{\rA}   &  \qw \\
 &  \pureghost{\Sigma}  &  \qw \poloFantasmaCn{\rA}   &  \measureD{\Tr} }   \end{aligned}~.
\]
Recalling that $\map D$ is a channel, and therefore  $\Tr_\rA  \map D  =   \Tr_\rB $  (proposition~\ref{prop:normalization}), we conclude that
\[
\begin{aligned}\Qcircuit @C=1em @R=.7em @!R {
& \prepareC{\rho}    & \qw \poloFantasmaCn{\rA} & \qw }\end{aligned}
~=\!\!\!\! \begin{aligned}\Qcircuit @C=1em @R=.7em @!R {
& \multiprepareC{1}{\Psi}    & \qw \poloFantasmaCn{\rA} & \qw \\  & \pureghost{\Psi}    & \qw \poloFantasmaCn{\rB}  &\measureD{\Tr} }\end{aligned}
~=\!\!\!\! \begin{aligned}\Qcircuit @C=1em @R=.7em @!R {
 &  \multiprepareC{1}{\Sigma}  &  \qw \poloFantasmaCn{\rA}   &  \qw \\
 &  \pureghost{\Sigma}  &  \qw \poloFantasmaCn{\rA}   &  \measureD{\Tr} }   \end{aligned}~.
\]
Hence,  the marginal of $ \Sigma $ on the first copy of system $\rA$ is equal to $ \rho $. By the same reasoning, we can prove that the marginal on the second copy of system $\rA$ is also equal to $\rho$.  Hence, $\Sigma$ is a symmetric purification of $\rho$.  
Since $ \rho $ is arbitrary, we conclude that every state has a symmetric purification, unique up to local reversible transformations.

Conversely, we now show that the second set of axioms implies the first.  To this purpose, we must show the validity of Local Exchangeability. 
  Clearly,   symmetric purifications are locally exchangeable:  indeed, if $\Psi$ is a symmetric purification one has  
\begin{align*}
   \begin{aligned}\Qcircuit @C=1em @R=.7em @!R { & \multiprepareC{1}{\Psi}    & \qw \poloFantasmaCn{\rA} &   \multigate{1}{  {\tt  SWAP}}   &  \qw  \poloFantasmaCn{\rA}  &  \qw   \\  & \pureghost{\Psi}    & \qw \poloFantasmaCn{\rA}   &   \ghost{  {\tt  SWAP}}   &    \qw \poloFantasmaCn{\rA} &\measureD{\Tr}  }   \end{aligned}    
   &=\!\!\!\!       \begin{aligned}\Qcircuit @C=1em @R=.7em @!R { & \prepareC{\rho}   &  \qw \poloFantasmaCn{\rA}  &  \qw  }  \end{aligned}  \\
   &  = \!\!\!\!        \begin{aligned}\Qcircuit @C=1em @R=.7em @!R { & \multiprepareC{1}{\Psi}    & \qw \poloFantasmaCn{\rA} &  \qw   \\  & \pureghost{\Psi}    & \qw \poloFantasmaCn{\rA}   & \measureD{\Tr}  }   \end{aligned}     ~ .  
\end{align*}
and, by the essential uniqueness of purification, 
\begin{align*}
   \begin{aligned}\Qcircuit @C=1em @R=.7em @!R { & \multiprepareC{1}{\Psi}    & \qw \poloFantasmaCn{\rA} &   \multigate{1}{  {\tt  SWAP}}   &  \qw  \poloFantasmaCn{\rA}  &  \qw   \\  & \pureghost{\Psi}    & \qw \poloFantasmaCn{\rA}   &   \ghost{  {\tt  SWAP}}   &    \qw \poloFantasmaCn{\rA} &\qw }   \end{aligned}   
 ~  =\!\!\!\!        \begin{aligned}\Qcircuit @C=1em @R=.7em @!R { & \multiprepareC{1}{\Psi}    & \qw \poloFantasmaCn{\rA} &  \qw  &\qw&\qw   \\  & \pureghost{\Psi}    & \qw \poloFantasmaCn{\rA}   & \gate{\map U}  &  \qw \poloFantasmaCn{\rA}    &\qw  }   \end{aligned}     ~ .  
\end{align*}
for some reversible channel $\map U$.   Since all purifications of $\rho$ are  equivalent to  $\Psi$ under local operations and since $\Psi$ is locally exchangeable, we conclude that every purification of $\rho$ is locally exchangeable [by the same argument used in Eq.~\eqref{aaa}]. This proves Local Exchangeability. \qed 

\section{Proof of proposition~\ref{prop:marginals}}\label{app:marginals}  

Let $\Phi  \in  \Pur\St \left(\rA\otimes \rA\right)$ be a symmetric purification of $\rho_\rA$ and let $\alpha$ ($\beta$)  be a fixed, but otherwise arbitrary, pure state of $\rA$  ($\rB$).  By the uniqueness of purification, there must exist a reversible transformation $\map U$ such that 
\[
\begin{aligned}\Qcircuit @C=1em @R=.7em @!R {    
& \multiprepareC{1}{\Psi}  &   \qw \poloFantasmaCn{\rA}       &  \qw    \\ 
&\pureghost{\Psi} &  \qw \poloFantasmaCn{\rB}  &  \qw   \\
    & \prepareC{\alpha}   &  \qw \poloFantasmaCn{\rA}  & \qw }     
\end{aligned} ~ 
= \!\!\!\!
\begin{aligned}\Qcircuit @C=1em @R=.7em @!R {    
& \multiprepareC{1}{\Phi}  &   \qw \poloFantasmaCn{\rA}       &  \qw    &  \qw  &  \qw    \\
&\pureghost{\Phi} &  \qw \poloFantasmaCn{\rA}  &\multigate{1}{\map U}  &  \qw \poloFantasmaCn{\rB}  &  \qw   \\
    & \prepareC{\beta}   &  \qw \poloFantasmaCn{\rB}  &  \ghost{\map U}  &  \qw  \poloFantasmaCn{\rA}   & \qw }     
\end{aligned} ~ .
\]
Discarding the first copy of system $\rA$ and using the fact that $\Phi$ is a symmetric purification we obtain the desired result
\[
\begin{aligned}\Qcircuit @C=1em @R=.7em @!R {    & \prepareC{\rho_\rB}   &  \qw \poloFantasmaCn{\rB}  &  \qw   \\
    & \prepareC{\alpha}   &  \qw \poloFantasmaCn{\rA}  &  \qw }     
\end{aligned}~
= \!\!\!\!
\begin{aligned}\Qcircuit @C=1em @R=.7em @!R {    & \prepareC{\rho_\rA}   &  \qw \poloFantasmaCn{\rA}  &\multigate{1}{\map U}  &  \qw \poloFantasmaCn{\rB}  &  \qw   \\
    & \prepareC{\beta}   &  \qw \poloFantasmaCn{\rB}  &  \ghost{\map U}  &  \qw  \poloFantasmaCn{\rA}   & \qw }     
\end{aligned}  ~ .
\]
      \qed

\bibliographystyle{apsrev4-1}
\bibliography{bibliography_Lo-Popescu}

\end{document}